# Engineering Quantum Wire States for Atom Scale Circuitry

Max Yuan[1], Lucian Livadaru[1], Roshan Achal[2], Jason Pitters[3], Furkan Altincicek[1], Robert Wolkow[1,2]

[1]Department of Physics, University of Alberta, Edmonton, Alberta, T6G 2E1, Canada
[2]Quantum Silicon Inc., Edmonton, Alberta, T6G 2M9, Canada
[3]National Research Council of Canada, Edmonton, Alberta, T6G 2M9, Canada

Corresponding email: rwolkow@ualberta.ca

**Abstract**
Recent advances in hydrogen lithography on silicon surfaces now enable the fabrication of complex and error-free atom-scale circuitry. The structure of atomic wires, the most basic and common circuit elements, plays a crucial role at this scale, as the exact position of each atom matters. As such, the characterization of atomic wire geometries is critical for identifying the most effective configurations. In this study, we employed low-temperature (4.5 K) scanning tunneling microscopy (STM) and spectroscopy (STS) to systematically fabricate and characterize six planar wire configurations made up of silicon dangling bonds (DBs) on the H-Si(100) surface. Crucially, the characterization was performed at the same location and under identical tip conditions, thereby eliminating artifacts due to the local environment to reveal true electronic differences among the line configurations. By performing dI/dV line spectroscopy on each wire, we reveal their local density of states (LDOS) and demonstrate how small variations in wire geometry affect orbital hybridization and induce the emergence of new electronic states. Complementarily, we employ density functional theory (DFT) and non-equilibrium Green's functions to compute the LDOS and evaluate transmission coefficients for the most promising wire geometries. Our results indicate that dimer and wider wires exhibit multiple discrete mid-gap electronic states which could be exploited for signal transport or as custom quantum dots. Furthermore, wider wires benefit from additional current pathways and exhibit increased transmission, while also demonstrating enhanced immunity to hydrogen defects.

## 1. Introduction

Semiconductor technologies are rapidly approaching the limits of conventional device designs. Although incremental progress is still being made in reducing node sizes and increasing transistor densities, the failure of power density scaling has constrained further performance improvements in complementary metal-oxide semiconductor (CMOS) devices.[1] However, the plateau of CMOS performance does not imply that all forms of silicon circuitry are similarly limited. One promising alternative employs patterned silicon DBs on an otherwise hydrogen-terminated silicon surface to realize a faster, lower-power, all-silicon approach.[2–5] While the benefits of atomic circuitry such as miniaturization and power efficiency are undeniable,[6] operating at atomic scales also introduces challenges, as electronic and quantum effects become increasingly prominent.[7,8] Understanding the behavior of individual atomic components is therefore essential for optimizing atom-scale beyond-CMOS devices.

An essential element for continued miniaturization is the atomic wire, which uses in-gap states to transmit signals across semiconducting surfaces, bias circuit elements, or function as extended quantum dots. Conventional CMOS devices rely on metallic interconnects fabricated using resist masks and metal deposition (atomic layer or physical vapor deposition).[9] However, such methods are unfeasible at the atomic scale due to the lack of atomic precision in the metal deposition process. Various atomic wire candidates have been proposed. Current atom-scale devices have predominantly used embedded dopant wires formed through a multi-step process including atomic lithography, dopant deposition, and silicon overgrowth.[10–15] While functional for some applications, encapsulated dopant wires have a broad spatial extend that is incommensurate with the closely spaced inputs and outputs of atomic circuitry. Lines with highly spatially confined states are required in such cases.

Silicon DB wires are an approach with significant potential to fulfill this need. Recent progress in hydrogen lithography (HL) has achieved error-free atomic fabrication through improvements in precision and error correction.[16,17] In the HL process, an ultrasharp STM tip selectively removes hydrogen atoms from the H-Si surface by injecting current at the target site, thereby exposing the unsatisfied DB orbital of the underlying silicon atom.[18–20] Conversely, errant DBs can be erased by functionalizing the tip with hydrogen and closely approaching the reactive orbital,[17] or by reacting DBs with $H_2$ molecules.[21] These DBs are variance-free and can host zero, one, or two electrons in well-defined energy levels within the silicon bandgap.[22,23] Additionally, a high H diffusion barrier (1.4 eV) renders DB structures stable up to 200° C, enabling operation under practical conditions above cryogenic temperatures.[24,25]

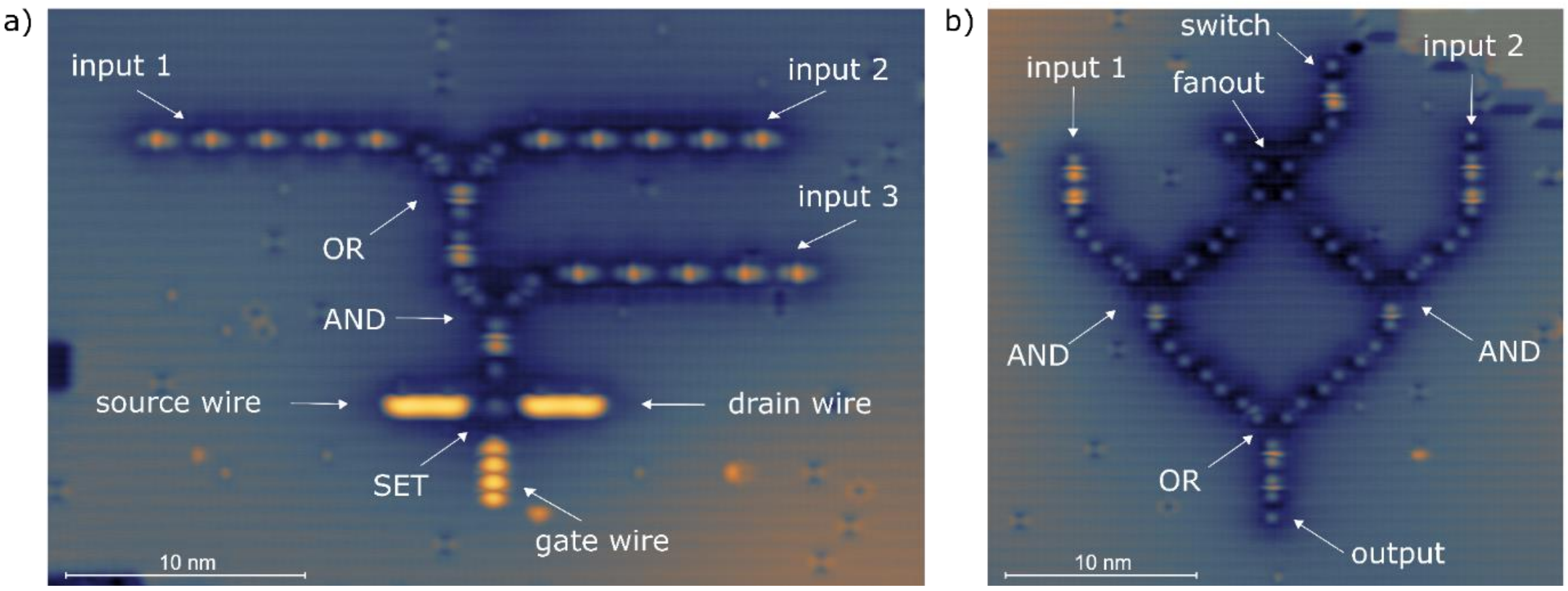

**Figure 1. Two atomic silicon integrated circuits are shown to demonstrate current capabilities to fabricate complex error-free silicon DB structures using STM-based HL on an H- terminated surface. (a) 1.3 V, 50 pA STM image of a three-input combination OR gate and AND gate with an atom-defined SET readout formed out of 69 DBs. (b) 1.3 V, 50 pA STM image of a two-input multiplexing DB circuit made of several gates and a fanout formed out of 56 DBs. Both circuit layouts were made using gate designs from binary atomic silicon logic [5] and were based on modeling and simulation software for silicon DB logic systems.[26]**

Recently, HL on H-Si has emerged as a promising platform for atomic circuitry. Demonstrations include logic elements, rewritable memory, controllable chemical reaction sites, and quantum devices.[5,8,21,26–30] Advances in precision, error correction, and lithography speed now allow for the manufacturing and study of high-complexity atomic circuits with zero errors. For example, Figure 1 presents two prototype circuit designs: one features cascading OR and AND logic gates with a single-electron transistor (SET) for readout, and the other shows a two-input multiplexing circuit. It is the aim of this and subsequent works to devise means to wire-up and actuate such circuits. Both circuits were designed using binary atomic silicon logic with SiQAD modeling software,[5,26] which calculates the ground state charge distributions to optimize DB layouts for each application. Using SiQAD,[26] DB ground state charge simulations were conducted for each circuit in Figure 1 to show their simulated functionality. The results are shown in Figures S1 and S2. Recent progress in silicon DB logic system modeling, automation, and simulation has also enabled large logic networks to be translated into atom-precise DB layouts, paving the way for scalable DB circuitry.[30–43] Although several obstacles remain, the present work focuses on developing conductive atom-defined wires composed of mid-gap states as exemplified by the source, drain, and gate wires in Figure 1a. Future studies will address additional challenges, including hermetic encapsulation (to prevent ambient gaseous molecules from annihilating DBs) and macro-to-atom connections for out-of-vacuum operation.[8]

The electronic behavior of these wires is complex, driven by factors such as the charge of individual DBs, orbital overlap, hybridization, local electrostatics, doping, and inherent instabilities.[44–51] Some of these can be modified by altering the atomic positions of DBs, resulting in different wire geometries. Various planar DB wire geometries have been proposed and studied.[44–51] In a theoretical study, Englund et al. investigated four wire types (straight, zigzag, dimer, and perpendicular) using DFT and transport calculations.[46] Their findings indicated that most of those structures are prone to structural distortions and magnetic ground states that hinder transport, whereas dimer wires appeared the most stable and conductive. In another theoretical work, Kepenekian et al. reported that dimer wires and zig zag wires are the least sensitive to instabilities, and that the zig-zag configuration offers enhanced conduction at low biases and reduced leakage to bulk states.[51]

While these wires have mostly been analyzed theoretically, some wire variations have also been studied experimentally. Previous research has reported both Peierls and Jahn-Teller distortions in single atom-wide DB lines.[19,52] Croshaw et al. used non-contact AFM to study single atom-wide DB wires of various lengths and confirmed that these wires adopt an ionic ground state with alternating positive and negative charges, as well as the existence of excited state configurations at higher voltages.[44] Altincicek et al. investigated silicon dimer wires on highly doped p-type silicon and found that each additional dimer contributes distinct filled and empty states to the wire.[45] Naydenov et al. studied single atom-wide, dimer, and wider wires using STS at 77 K, revealing “quantum well” states for dimer wires and evidence of metallic behavior.[50] Other linear DB structures studied included H-capped silicon lattice sites between DBs and as a result are expected to have greatly reduced conduction relative to the wires studied here.[53]

In this work, we expand upon previous studies by using STS and DFT to measure and compare the LDOS of various DB wire geometries at 4.5 K. To determine the effects of wire geometry on electronic properties, we first fabricated a single DB, then sequentially created six DB wires of different geometries at the exact same surface location, erasing previous structures when necessary, thereby eliminating artifacts due to the local environment to reveal true electronic differences among the line configurations. All wires were 4.6 nm long (12 surface lattice constants along the dimer row direction), though their orbital extents slightly exceed this length. Figure 2 displays ball-and-stick models along with filled and empty state STM images of the studied structures. Constant height dI/dV spectroscopy reveals the LDOS for each wire configuration and identifies discrete electronic states in the silicon bandgap which are crucial for conduction without bulk leakage. Additionally, ab initio calculations provide further insight and enable the computation of transmission coefficients for the most promising geometries. This systematic approach systematically elucidates how DB geometry affects electronic behavior and identifies prime wire candidates for DB-based circuitry.

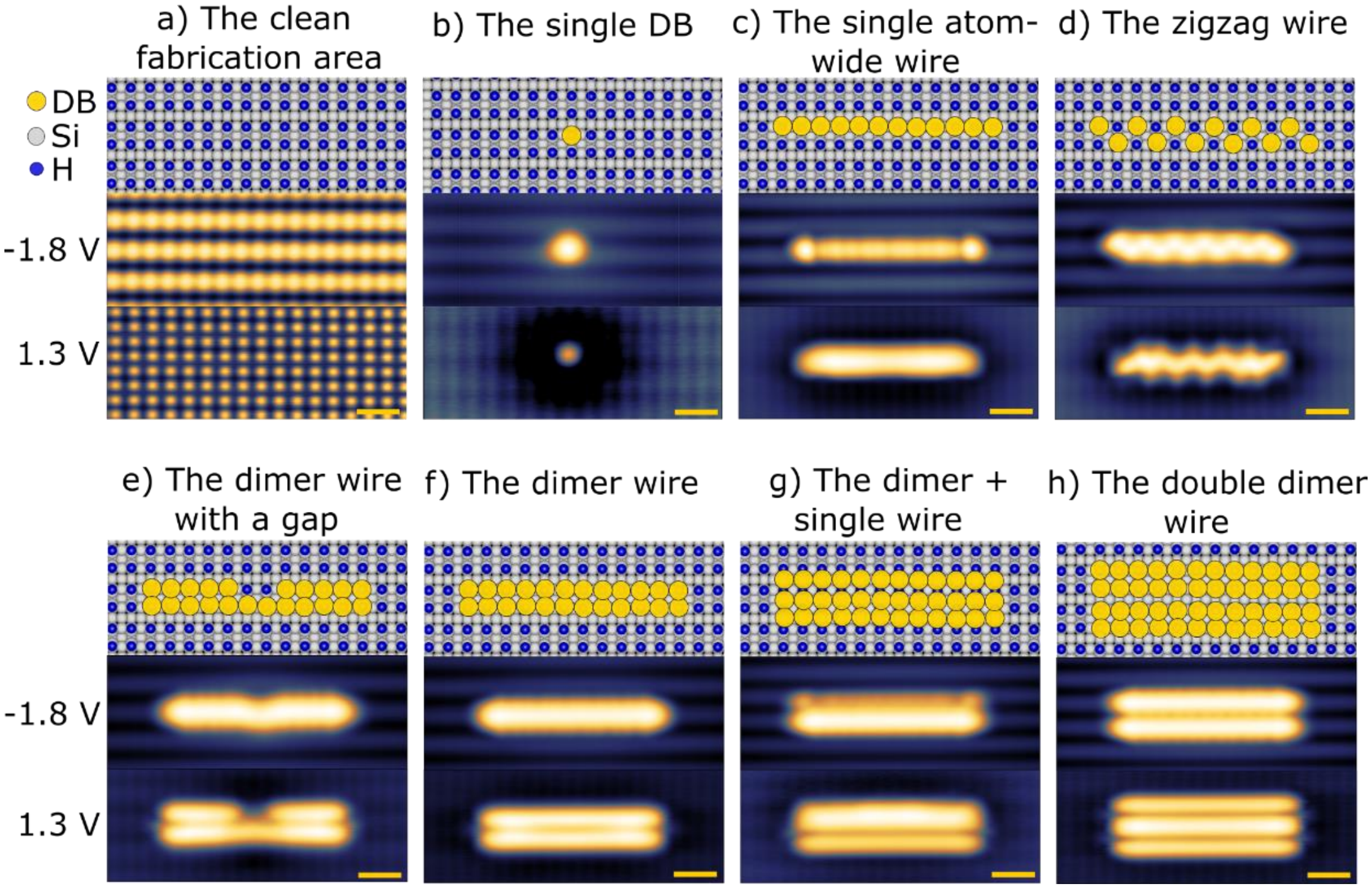

**Figure 2. (a-h) Ball and stick diagrams, and STM images ($V_{sample}$= -1.8 V and 1.3 V, I = 50 pA), showing the same crystalline area with the sequence of structures that were fabricated and measured starting with a single DB and ending with a 12 long double dimer wire. 1 nm bars are included in the bottom right corner of the images for scale.**

## 2. Results and Discussion

### 2.1 The Single DB

The 2x1 H-Si(100) surface is defined by a repeating surface unit consisting of covalently bonded pairs of silicon atoms (dimers) with each atom capped by an H atom (i.e. a 2x1 dimerized structure). To begin, a 25 nm × 15 nm area of H-Si(100) free of defects and charge centers was selected for fabrication (partially shown in Figure 2 and fully in Figure S3). A 1250 °C crystal flashing procedure was done during sample preparation to produce a near-surface dopant depletion region that effectively decouples the DB wires from the donor band.[54] This decoupling is crucial as samples with higher near-surface dopant densities may exhibit spectral peaks that mask the electronic signatures derived from DBs and their ensembles (Fig. S4).[23,54–56]

Before constructing the first wire, a single DB was created and analyzed to serve as a reference. Figures 3a and 3b present I-V and dI/dV measurements performed on the DB and at a nearby H-Si location. For a given bias V, current is proportional to the integral of the LDOS between the tip and the sample Fermi levels, while the dI/dV spectrum is proportional to the LDOS at each bias, allowing individual states to be resolved.[57]

The I-V spectroscopy over the H-Si surface exhibits typical semiconductor behavior: a clear band gap with gradual current increases at the valence and conduction band edges.[57] Based on the current onsets, the apparent band edges were identified at –1.60 V and 0.98 V (defined where the current exceeds 1 pA), yielding an experimental band gap of 2.58 eV. This apparent gap is larger than the known silicon band gap (1.17 eV at 4.5 K), a discrepancy attributed to tip field–induced band bending (TIBB).[58]

In the dI/dV plot of Figure 3b, a prominent DB peak is centered at –1.50 V, which has been previously associated with a DB charge transition.[55] There also exists a positive bias charge transition peak, however it is located beyond the range of this spectroscopy. Figures 3c–f present STM images of the DB at different bias values, exhibiting various charge states. The charge state of the DB depends on the balance of electron filling and emptying rates, which can be altered by changing the tip bias and height.[22,55,59] To illustrate this, imaging the DB to the left of the charge transition peak (-1.80 V, Figure 3c) renders it negative, whereas imaging it to the right of the peak (-1.30 V, Figure 3d) renders it positive. Additionally, imaging the DB at the same bias as the transition peak shows an on average neutral DB since the emptying rate to the tip matches the filling rate from sample (see Figure S5).[22,55,59] In the absence of the tip, isolated DBs on this surface are inherently negative due to the degenerate n-type doping of the crystal.[5,44] It is important to note that, even at zero bias, the W tip/n-type silicon contact potential of approximately 1 eV can render a DB neutral if the tip is sufficiently close.

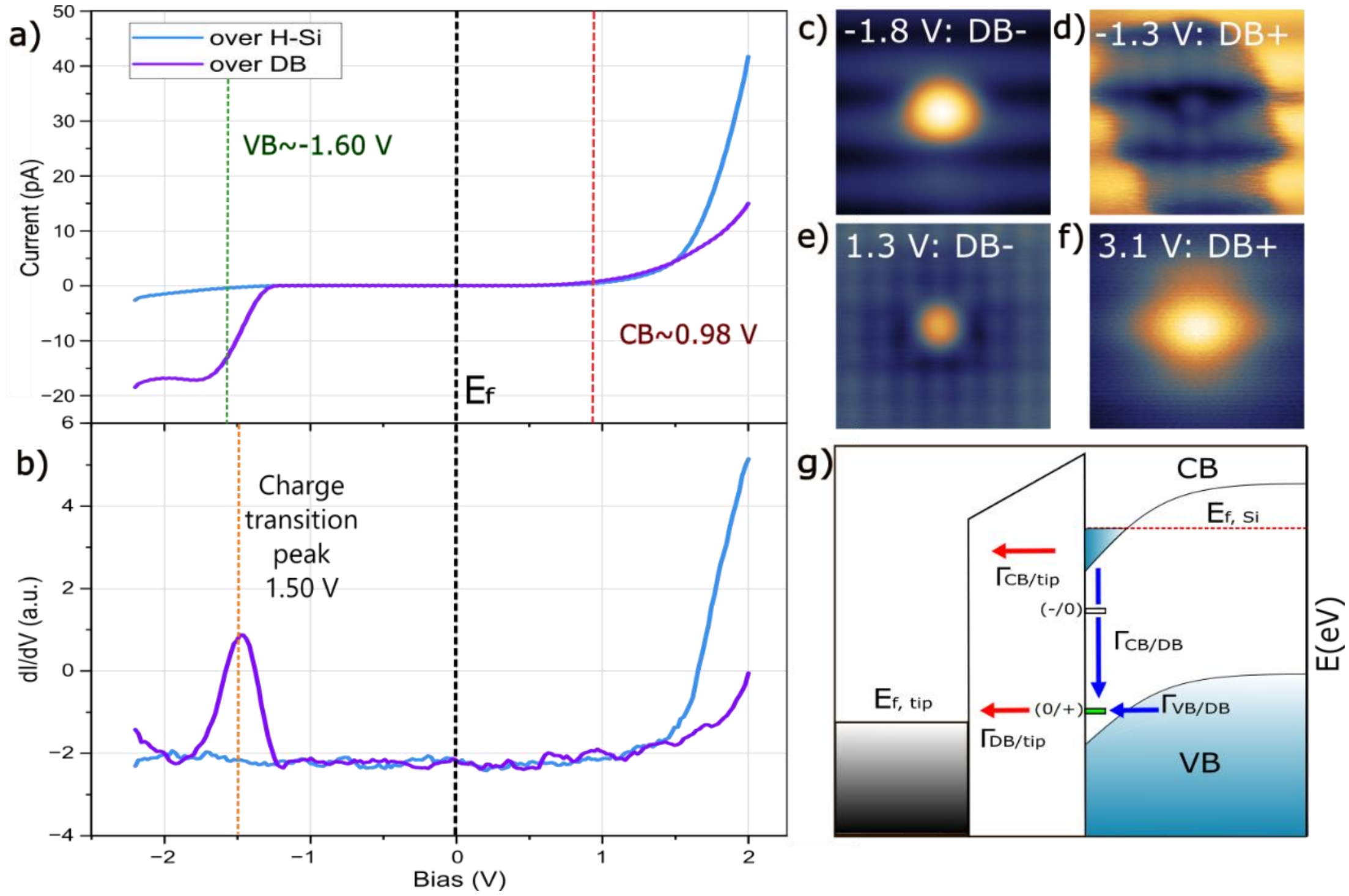

**Figure 3. (a) I-V curves measured over H-Si and over a DB at a setpoint 2 V, 50 pA over H-Si. Current onsets over H-Si are used to determine the apparent valence and conduction band edges for the crystal. (b) dI/dV spectroscopy curves done over the same species. (c-f) STM images of a single DB at specified bias values, with charge state indicated for each case [54]. (g) Qualitative band diagram in a negative bias case showing the tip, DB, and sample energy levels, as well as the tunneling pathways and competing rates for electrons in the DB levels. Here, a biased tip is extracting electrons from the lower (0/+) level which is being supplied from bulk VB levels by tunneling and the bulk CB levels by inelastic electron capture mediated by the (-/0) level.**

Figure 3g schematically illustrates the current pathways in a negative bias case: electrons transit from occupied bulk states into the DB by two different pathways. Here, tip-induced band bending aligns the (0/+) DB level in resonance with the valence band, enabling electron filling from both the valence band via tunneling to the (0/+) DB level and from the conduction band via inelastic capture. From the DB, electrons tunnel across the vacuum barrier to unoccupied tip states. The time-average charge of the DB at a given bias is then determined by the exact balance of electron filling rate from the bulk and emptying rate to the tip.[55,56]

The DB's charge state also influences the local lattice configuration. Neutral DBs are nearly unbuckled (dimer bond angle approximately –1.2° w.r.t. the surface) with their silicon atom positioned only slightly lower than a normal H-terminated silicon atom. In contrast, negatively charged DBs (which hosts two electrons) exhibit an upward buckling of about 8°, lifting the silicon atom by roughly 30 pm and adopting a more $sp^3$-like hybridization.[44,60] Conversely, when the DB is positive (zero electrons), the silicon atom buckles downward by approximately 10°, lowering the atom by about 40 pm and resulting in a more $sp^2$-like hybridization.[44,60]

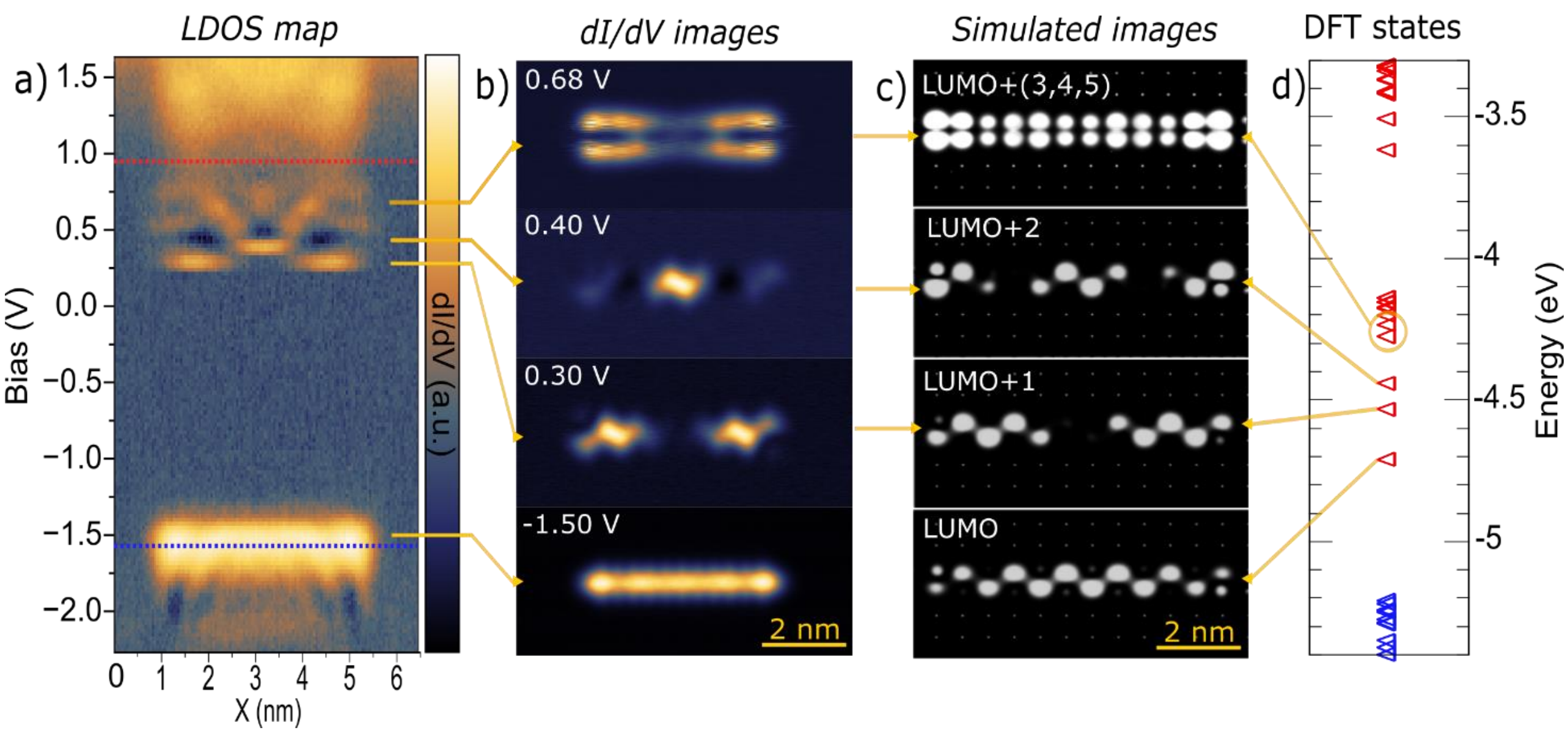

**Figure 4. (a) dI/dV LDOS map taken over the 12 long dimer wire using a 1.8 V 50 pA tip set point over H-Si. Apparent band edges are shown in blue (VB) and red (CB). (b) Constant height dI/dV images of the wire from the same setpoint. Energies were chosen to correspond to states shown in (a). (c) Simulated LDOS for the first several states in the band gap. (d) A calculated energy level diagram for the dimer wire showing the manifold of filled (blue) and empty (red) states for the wire, analogous to the experimental data in (a).**

### 2.2 DB Wires

Following our single DB characterization, additional DBs were subsequently added or removed to sequentially form six distinct atomic wire geometries, all fabricated and characterized in the same location. This procedure minimizes variations from local electrostatic inhomogeneities that would otherwise arise if the wires were produced in different regions (see Figure S6 for examples).[61] Equally important, each wire was analyzed under equivalent tip apex conditions, which can also significantly affect spectral measurements (Figure S7).[62–64] Tip characterization was accomplished by repeatedly collecting I-V spectroscopy data over a reference DB situated sufficiently away from the wire between each measurement session (Figure S8). If any variations were observed, the tip was re-conditioned until the original reference DB spectrum was reliably reproduced. Furthermore, the same setpoint parameters (1.8 V, 50 pA over H-Si) were maintained throughout the experiment. Altogether, these consistencies ensure that the wire spectra for all wire geometries are directly comparable.

As with the single DB, dI/dV spectroscopy was used to probe the LDOS of the various wire geometries. Due to the one-dimensional nature of the wires, multiple spectra (100 per wire) were acquired along their lengths and combined to create two-dimensional LDOS maps (see the Methodology section for details).

Measurements and DFT calculations for the dimer wire are presented in Figure 4. This structure was fabricated by creating 24 DBs, thus exposing 12 bare dimers. Dimer wires are particularly promising because of the interaction of $\pi$ bonds between dimerized atoms which enables transport.[46,50,65,66] Moreover, such wires are reputed to be remarkably insensitive to geometric and magnetic distortions.[46,47] Figure 4a shows the LDOS map for the dimer wire, with a highest occupied molecular orbital (HOMO) detected at –1.50 V and a lowest unoccupied molecular orbital (LUMO) at 0.30 V. Overlaying the experimentally measured band edges on the spectroscopy confirms the presence of several unoccupied states within the bandgap, a critical feature that permits signal transmission through the wire without substantial leakage to the bulk Si states. The alternating bright and dark features correspond to the peaks and troughs of the electron wavefunction extending along the wire, indicating effective hybridization and delocalization.[50] Moreover, the discrete energy levels of these states suggest they could be individually addressed, potentially allowing their use as custom-engineered electronic quantum dots.[67]

Constant-height dI/dV scanning was also performed to image the wire at specific energies (Figure 4b). These images closely resemble the corresponding LDOS maps. Similar to the single DB, the underlying Si atoms in the dimer wire undergo lattice distortions buckling upward ($sp^3$) or downward ($sp^2$) depending on the DB's charge.[60] For example, the LUMO (at 0.3 V) and LUMO+1 (at 0.4 V) images exhibit a zigzag pattern, indicating that dynamic dimer buckling (present in other conditions) is frozen at these low biases.[68] This buckled arrangement suggests a ground state composed of alternating charged bare dimers [(-+) and (+–)]. At higher biases, the wire appears vertically symmetric, averaging over the various buckled geometrical configurations (as seen in the top image of Fig. 4b)[69]; this is why states in the spectroscopy above 0.5 V appear less distinct than the lower-energy, geometrically frozen states.

DFT calculations were undertaken to simulate and visualize the electronic states of the dimer wire (the simulation methods are detailed in the Methodologies section). Figure 4d shows the energy levels for the 12-dimer (24-DB) wire. Unlike the two discrete states observed for the single DB in Figure 2g, the dimer wire exhibits a broad manifold of filled (blue) and empty (red) states, many within the bandgap. The calculated LDOS for the first few empty states is illustrated in Figure 4c. The LDOS maps for the second, third, and fourth calculated states in the band gap closely match the corresponding dI/dV images in terms of the number and shape of nodal features. Note that the topmost theoretical LDOS map is derived as a sum over three closely spaced states (indicated by the yellow circle) that cannot be individually energy-resolved due to experimental resolution of approximately 100 meV.

The first theoretical LUMO state in the band gap exhibits low signal and is not resolved in this experiment. However, separate measurements conducted with different tip apex conditions (outside this dataset) have resolved it (see Figure S7). The empty states display an increasing number of nodes with energy.[50] The energy spacing between consecutive states differs in experiment and theory due to band bending at the surface, caused by two factors: (i) dopant depletion in the subsurface region, which shifts the observed surface Fermi level relative to the bulk, and (ii) tip-induced band bending during STS, which varies with the applied bias. Furthermore, the DFT-calculated energy levels exhibit some artifacts due to the approximations inherent in the model and the exchange-correlation functional used (PBE in our case). Consequently, the gap levels (relative to the VBM) are artificially "stretched out," with a model bandgap of 1.54 eV versus an expected 1.17 eV in reality (versus 2.58 eV apparent experimental bandgap).

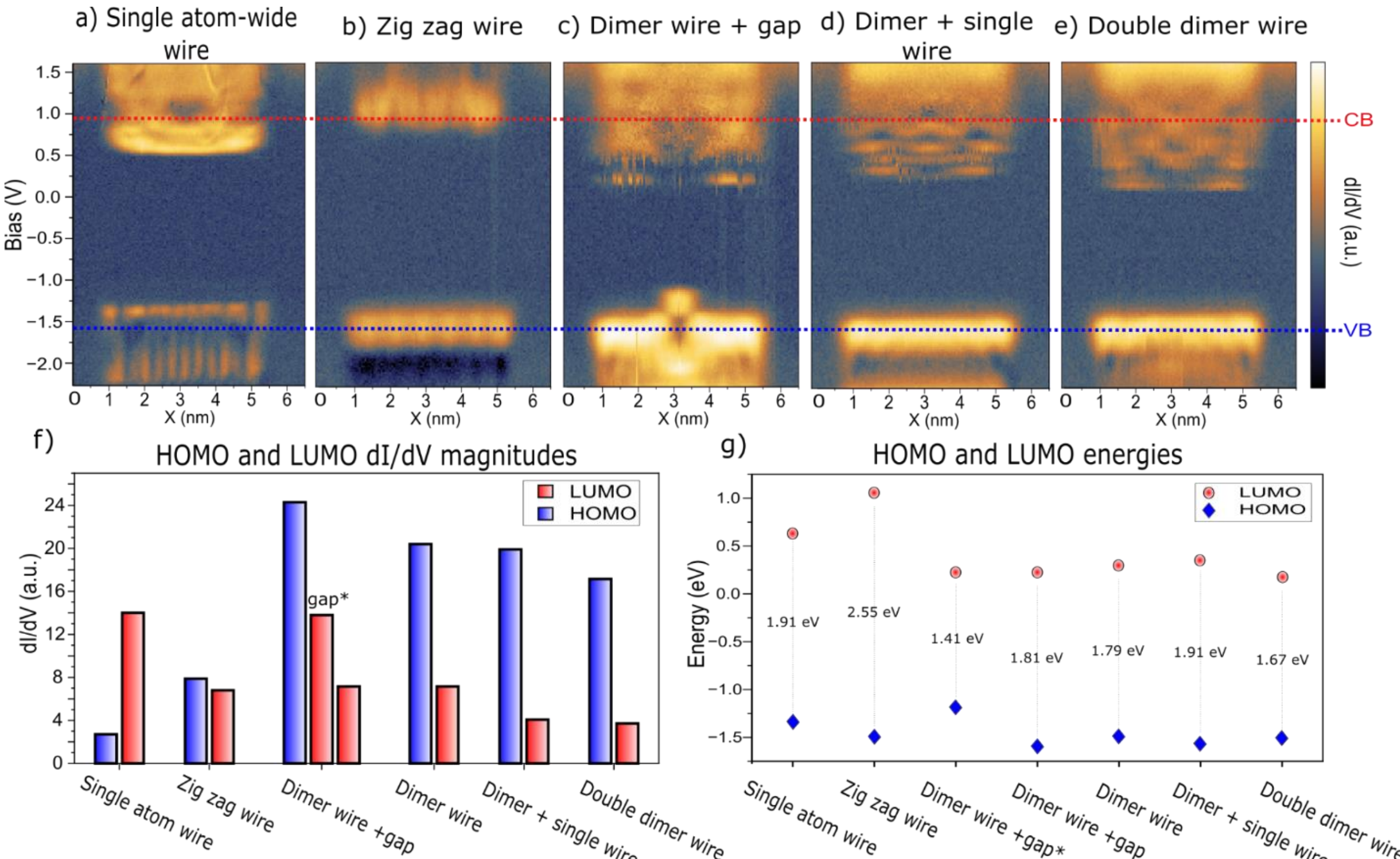

**Figure 5. (a-e) dI/dV LDOS maps for the 12 long single atom-wide wire, zig-zag wire, dimer + 2H gap wire, dimer + single atom-wide wire, and double dimer wire. Overlaid are the apparent band edges which were extracted from the H-Si spectra in Fig. 3. (f) The magnitude of the HOMO and LUMO dI/dV signals for each wire (not distinguishable in the spectra due to logarithmic processing, see Fig. S15 for the unprocessed data). (g) The HOMO and LUMO energies as well as the apparent band gap for each wire (extracted using the peaks of each state). The dimer wire + gap* shows the values directly over the H defect.**

Figure 5 displays the side-by-side comparison of the measured spectra for all the other wire geometries, alongside their HOMO and LUMO state information (constant-height dI/dV images are provided in the Supplementary Information). The simplest variant is the straight 12-DB single atom-wide wire (Figure 5a), formed by adding 11 DBs to the initial DB characterized in Figure 3. The negative-bias portion of this spectrum resembles the spectra of the single isolated DB, with a homo peak at -1.35 V (instead of -1.50 V) that may be similarly attributed to a charge transition within the wire.[55] The LUMO appears below the conduction band at 0.60 V and is separated from the HOMO by 1.91 V. Although its positive bias states are not sharply defined, they may facilitate signal transmission. However, single atom-wide wires are reported to be more reactive with ambient molecules than wider geometries,[21] and earlier calculations indicate that they are more susceptible to geometric and magnetic distortions that can inhibit conductivity.[46]

For the next wire structure, half of the DBs in the single atom-wide wire (Figure 2c and Figure 5a) were erased and then replaced on the opposite sites of the same dimers, resulting in the zig-zag wire (Figure 2d and Figure 5b). Although DFT transport calculations by Kepenekian et al. predicted that the zig-zag wire is resistant to instabilities and may offer good conduction at low voltages,[51] the corresponding LDOS map (Figure 5b) reveals a scarcity of in-gap states available for conduction and exhibits the largest apparent bandgap (2.55 eV) among the tested wires. This wider gap is expected due to the increased distance and weaker interaction between nearest-neighbor DBs compared to the straight single atom-wide wire.

Next, the zig-zag configuration was modified into a gapped dimer wire (Figures 2e and 5c) by adding DBs and intentionally leaving a two-hydrogen gap on one side. The presence of this defect reduces the LDOS in the central region at positive bias and causes a new split state to emerge at negative bias. This splitting appears consistently in both the experimental data and the DFT calculations (Figure S12) for this DB geometry. Despite the apparent break in the LDOS, the wire wavefunction remains evanescently connected in that middle region. Interestingly, at a positive bias of 0.23 V, the "streakiness" in the LUMO states indicates a larger scale switching, as the LDOS appears to toggle between two degenerate configurations located on either side of the defect (as detailed in Figure S12a). This is also supported by our DTF results. The degree to which the hydrogen defect impairs conductivity will be analyzed via NEGF-DFT transmission calculations in the theory section.

Building on the dimer wire (Figure 4), wider wire configurations were created, including the "dimer + single atom-wide wire" (Figures 2g and 5d) and the "double dimer wire" (Figures 2h and 5e). For both, spectroscopies were taken at the center of the dimer row (additional measurements offset from the row center are shown in Figure S14). Widening the wire has a notable effect on the filled states: for instance, the double dimer wire shows increased LDOS at –1.90 V compared to both the dimer and the dimer + single wire configurations. This enhancement indicates lateral coupling between the two neighboring dimer wire components. This is supported in Figure S14, where the spectroscopy taken over the dimer row shows the same empty states appearing between 0.20 V and 0.80 V as in the spectroscopy taken between dimer rows. Such observations indicate that signal propagation between dimer rows is feasible in these wider wires. Furthermore, the empty states of the double dimer wire become more complex due to additional hybridization between DBs, and they extend lower into the Si bandgap, reducing the HOMO–LUMO gap to 1.67 eV. Compared to previous wire geometries, the larger manifold of discrete bandgap levels with strong delocalization indicates enhanced electron transport characteristics with low leakage to bulk for these wider wire types.

The double dimer wire also exhibits improved mechanical stability, as evidenced by reduced dynamic dimer buckling during measurements. This stabilization expands the voltage range over which the wire remains in a fixed buckling configuration, resulting in more sharply defined bandgap states with an increasing number of nodes as bias increases. In the dI/dV images (Figure S13), the buckling in the double dimer is static up to 0.68 V, in contrast with dynamic buckling in the dimer wire at the same bias (Figure 4b). This points to an added advantage of the wider wires. Since collective current channels maintain conductivity even in the presence of such defects their sensitivity to missing or misplaced DBs is reduced.

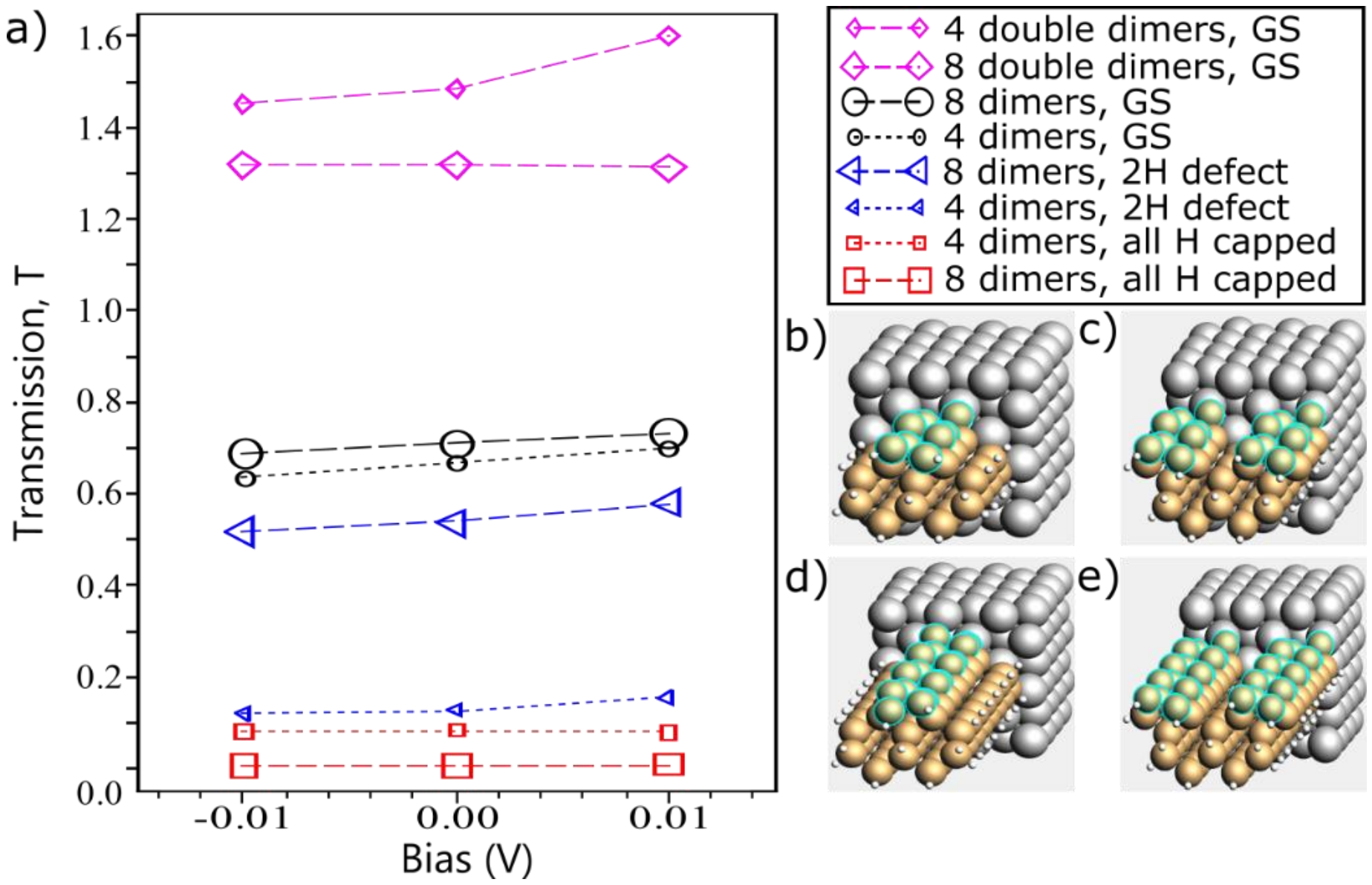

**Figure 6. Quantum transport calculation through 4- and 8-long dimer wires on Si nanoclusters sandwiched between metal electrodes, predicted using the NEGF-DFT method in the limit of low to zero bias. (a) The compiled values of the transmission function as a function of bias for various wire lengths and types, indicated in the legend. GS stands for the "ground state" configuration of the wire. The lines connecting the symbols are just a visualization aid. The remaining panels indicate the ground-state geometries used in the**

**calculations of four wire types: (b) 4-long single dimer wire; (c) 4-long double dimer wire; (d) 8-long single-dimer wire; (e) 8-long double dimer wire. Si atoms are depicted in yellow, H atoms in white, Ag atoms in gray, with the Si atoms hosting DBs being highlighted in light green. Only the far-side electrode is shown for clarity.**

We also performed quantum transport calculations using non-equilibrium Green's functions coupled with DFT (NEGF-DFT) to compare the conduction properties for the more promising wire candidates, the dimer and double dimer wires. This enables us to obtain theoretical predictions of quantum transmission (*T*) in the low-bias limit for relatively short wires of different types, using the method detailed in the *Methodologies* section. Each wire type was sandwiched between two identical metal contacts acting as source and drain. Three principal aspects were investigated: *(i)* whether the transport is ballistic; *(ii)* The scaling of current with increasing wire width; and *(iii)* the impact of specific defects on transmission.

Figure 6a shows the transmission function *T* close to zero bias for the various wires (wire types are indicated in the legend). Each wire geometry is fixed to the ground state (previously determined by a DFT geometry optimization calculation). All defect-free wires exhibit substantial zero-bias transmission (same per each spin orientation). Notably, when comparing the single dimer wires of lengths 4 and 8 (black symbols), the zero-bias transmission remains largely unchanged (around 0.7 and slightly higher for the longer wire) indicating ballistic conductance for this wire type. The transmission does not reach unity because, unlike previous NEGF studies in the literature where the system consisted only of silicon,[46,49,70] our system includes metal electrodes and thus metal-silicon interfaces, yielding a finite reflection coefficient, and thus diminishing the total transmission.

To determine whether electron transmission occurs along the wire, through the surrounding silicon cluster, or even via the vacuum, we selectively "capped" specific DBs with hydrogen atoms to remove their contributions to the electronic wavefunction. When all DBs in a wire were capped, leaving only a plain H-Si surface, the transmission *T* dropped dramatically. A small residual transmission (0.08 for the 4-long wire, versus 0.005 for the 8-long wire) can be attributed to residual through-vacuum tunneling, which decays exponentially with increased electrode separation.

To test the resiliency of transport to small defects in the wire structure, we introduced a defect by capping two DBs in the middle of the wire (similar to the defect in Figures 2e and 5c). The blue curves in Figure 6a show that transmission decreases for both wire lengths, with *T* for the 4-long wire dropping from 0.67 to 0.12 and for the 8-long wire from 0.71 to 0.54. Thus, a longer wire demonstrates greater resilience: the 8-long wire experiences only a 24% loss in zero-bias transmission compared to an 82% loss for the 4-long wire when a two-H defect is introduced. Although computational limitations prevented modeling longer wires, we expect this trend to persist, allowing minor imperfections in long wires without completely compromising conductivity.

Finally, we doubled the width of the defect-free dimer wires and obtained the results shown as pink symbols in Figure 6a. The transmission nearly doubles with the increased width. For example, at zero bias the transmission for a 4-long double-dimer wire is 1.48, more than twice that of a 4 single-dimer wire (0.67); for an 8-long double-dimer wire, *T* is 1.32, just short of twice that of an 8 single-dimer wire (0.71). Slight deviations from perfect linear scaling of *T* vs width are caused by the significant lateral interaction between the individual single dimer wires composing the double-dimer wire. Also note that, in building the model systems to compare the single- and double-dimer wires, a subtle technical issue arises: while the longitudinal (*z*-axis) spacing at the Si/Ag contact was maintained in all cases, it is impossible to impose the same consistent alignment in the lateral (*xy)* plane between electrode atoms and the end Si atoms of both dimer wires due to the mismatch in the crystal lattice constants of the two materials.

Our transport results predicting good transmission for DB wires around zero bias are largely consistent with previous studies.[46,49,70] Building upon these previous works, three significant additional factors included in this study are the inclusion of more realistic metal contacts, improved transmission in the presence of H defects, the analysis of wider double dimer wires, and the accompanying experimental data. Whereas earlier studies used highly doped silicon substrates as electrodes (thereby relying on narrow impurity, or donor, bands with widths of only tens of meV near the conduction band edge), our approach employs Si/Ag interfaces. The impurity-band model implies that conduction is very sensitive to any misalignment between the wire levels and the bias window. By contrast, our metallic leads provide broad, continuum states around the electrode Fermi level but introduce some scattering at the contacts. As a result, we do not expect the same extreme sensitivity to defects as reported in earlier work. For instance, while [[46]] documented a dramatic drop in zero-bias transmission upon the adsorption of a single H atom (capping a DB), our results in Figure 6a show that a significant fraction of the original transmission survives even when two H atoms are adsorbed. Similarly, whereas those studies found a zero-bias transmission of 1, we obtained approximately 0.7, a difference that can be attributed to the presence of our Si/Ag contacts rather than an all-silicon system found in the other studies.

### 2.3 Outlook

By combining our experimental spectroscopic data with the NEGF-DFT transmission results, we can now discern which wire geometries are most promising for circuitry and device applications. Table 1 summarizes these findings, along with ratings for their potential uses in transport or quantum dot applications.

| Wire type | Rating | Properties |
| --- | --- | --- |
| Single atom-wide wire | Poor | - Marginal in-gap states<br>- Reactive/unstable<br>- Poor transmission |
| Zig zag wire | Poor | - No in-gap states |
| Dimer Wire with 2H defect | Medium | - In-gap states<br>- Defect lowers transmission without eliminating it |
| Dimer wire | Good | - In-gap states<br>- Good transmission |
| Wider dimer wires | Best | - In-gap states<br>- Best transmission<br>- Most resilient to small defects and mechanical fluctuations |

**Table 1. Summary of the wire findings, rating their applicability for transport or quantum dot applications.**

The simplest geometry, the single atom-wide wire, exhibits only a marginal manifold of in-gap states close to CB edge. In transport, these are prone to leakage to bulk states. Prior calculations also suggest that such wires suffer from poor transport characteristics largely due to the formation of polarons when additional electrons are injected.[46] Moreover, single atom-wide wires are more reactive, as they tend to bond with stray $H_2$ molecules in the chamber, slowly erasing the wires even under vacuum conditions.[21] They also run the risk of DB diffusion/ migration across the dimer rows (despite a high diffusion barrier) under elevated temperatures or nearby electric fields, potentially destroying the intended structure.[25] These drawbacks render the single atom-wide wire impractical. Similarly, the zig-zag wire displays no discernable in-gap states, leading to leakage of any injected current to the bulk, and making it unsuitable for transport or quantum dot applications.

In contrast, the dimer wire geometry exhibits significant and discrete in-gap electronic states that extend about 700 mV into the apparent bandgap. These states arise from the interaction of π bonds forming multi-dimer collective states,[71] a behavior that implies significant delocalization. Our NEGF-DFT transport calculations indicate that the dimer wire exhibits an almost length-independent transmission coefficient of roughly 0.7 around zero-bias, an indicator of good transport properties. In addition, introducing a 2H-defect in the 8-long dimer wire only reduces transmission by 24%, confirming that these wires are more resilient to defects than previously assumed. Further stability is provided by their inertness toward diatomic molecules like $H_2$ and by eliminating the effects of DB diffusion within a dimer.[22] This inherent stability, along with their discrete mid-gap states, makes the dimer wire promising for both transport and quantum dot applications.

Wider wires share many of the dimer wires favorable characteristics, with the double dimer wire being especially notable. Its states extend deeper into the bandgap (with a LUMO at 150 mV, compared to 300 mV for the dimer wire), thus affording greater isolation from bulk levels, and a richer spectrum of in-gap states. NEGF-DFT calculations reveal that the double dimer wire averages a transmission coefficient of around 1.4, nearly double that of the single dimer wire due to the availability of multiple current channels. These extra channels also render wider wires more resilient to H defects, as alternative conduction pathways can bypass local disruptions. For applications requiring increased current, wider dimerized wires can be deployed. Looking ahead, future experiments including multiprobe STM measurements similar to those demonstrated for Ge(100)[72] and H-Si(100)[73] will directly assess transport along these wires.

Despite the progress made with individual components such as wires, logic gates, rewritable memory, and quantum dots, two major challenges remain for integrating complete devices: (1) preventing surface contamination when removing wafers from vacuum and (2) establishing reliable connections between macroscopic leads and atomistic circuitry. Proposed solutions for encapsulation include wafer capping techniques where H-Si samples are bonded to protect the surface during air exposure,[74] or the fabrication of vacuum-tight cavities through microfabrication methods.[75–77] For connecting atomic circuitry to macroscopic electrodes, metal or metal silicide deposition is the most straightforward method.[78–80] However, the out-of-plane geometries and interface junctions present challenges in coupling to DB wires. An alternative may be the use of dopant-implanted leads, which have proven capable of controlling DB structures and lie in-plane with the H-Si sample.[81,82] Future studies will address these issues as we progress toward practical DB devices.

## 3. Conclusion

In summary, we have demonstrated the fabrication of complex, error-free atomic circuitry for beyond-CMOS electronic devices, addressing an important challenge: the creation of atom-scale wires for signal transmission and circuit element biasing. We iteratively fabricated and characterized structures starting with a single Si DB and following with six different DB wire geometries, showcasing the development toward high-complexity atomic circuits exemplified by the prototypes in Figure 1. Crucially, all wires were measured under identical tip conditions and at the exact same crystal location to ensure direct and reliable comparison of their electronic properties. Using dI/dV spectroscopy, we extracted the LDOS for each geometry to evaluate their suitability for transport or quantum device applications. NEGF-DFT simulations further provided transmission coefficients and helped clarify the role of collective in-gap states in potential conduction pathways.

The single DB exhibits a prominent spectral feature at –1.5 V, identified as a charge transition. As this DB is extended into wires, additional features emerge due to DB interactions. Both single atom-wide and zig-zag configurations were found to have minimal discrete in-gap states, rendering them unsuitable for practical use. In contrast, dimerization produces discrete in-gap states attributable to π bonding, which in turn yield delocalized wire states, making them viable as practical atomic wires. Wider structures, such as the double dimer wire, inherit these favorable characteristics while showing improved transmission (due to additional current channels) and significant coupling between dimer rows. This may enable signal propagation even perpendicular to the dimer row direction, a valuable feature given the anisotropic nature of the H-Si (100) surface. Overall, our work identifies the dimer and wider dimer wires as strong candidates for both transport and quantum device applications.

## 4. Methodologies

**4.1 System** Experiments were conducted using a commercially available Scienta Omicron LT STM system operating at 4.5 K and controlled via Nanonis SPM electronics. The scan chamber was kept at ultra-high vacuum conditions with a pressure of $2.5\times10^{-11}$ Torr.

**4.2 Sample Preparation** The sample used in the experiments was highly arsenic doped (n-type) Si (100) with a resistivity of 0.003-0.005 Ohm·cm. The sample was degassed at 600° C in UHV overnight to reduce outgassing during subsequent flashing. Using resistive heating, the sample was flashed to 1250° C several times to remove the oxide and recrystallize the surface. The sample was then exposed to $10^{-6}$ Torr of cracked ultra-pure hydrogen gas. In the presence of the hydrogen gas, the sample was then flashed to 1250° C and then held at 330° C for two minutes resulting in a clean H-Si (100) sample.

**4.3 Tip Preparation** Poly-crystalline tungsten wire (0.25 mm diameter) was initially electrochemically etched in a 2 M NaOH solution resulting in a sharp, though not atomically sharp, apex. The tip was then brought into a custom-made field ion microscopy UHV chamber for in-situ sharpening. Helium gas was used as the imaging gas to visualize the apex, and nitrogen gas was used as the etching agent resulting in a single atom tip. See [[83,84]] for the complete FIM procedure.

**4.4 Hydrogen Lithography** DB fabrication was done from a starting bias of 1.3 V 50 pA using 50 ms pulses to raise the bias between 1.9 and 2.3 V. Most lithography was done in constant current mode, though some sections required constant height scanning to maintain the desired tip sample separation. Errant DBs were erased by functionalizing the tip with atomic hydrogen and lowering the tip approximately 500 pm from the 1.3 V 50 pA setpoint over the target DB. More details on the DB creation and erasure procedures are found in.[16,17]

**4.5 Spectroscopies** Each dI/dV LDOS map was obtained by combining 100 spectra along the lengths of the wire and processed using a custom python script. The log of the measured data was presented to enhance visibility of spectral features. The data is presented on a linear scale in Figure S15. dI/dV was calculated using the built in Nanonis lock-in amplifier, as opposed to calculated numerically. The lock-in was run at 700 Hz with a modulation voltage of 25 mV. The sequence of spectroscopies was conducted with the tip at a constant height and runs were limited to around 15 minutes to reduce the effect of tip drift. All wire measurements used a common tip setpoint of 1.8 V, 50 pA over the same H-Si atom.

**4.6 Electronic Structure Calculations by DFT** Geometry optimization calculations on multiple DB wire types were performed by DFT using the AMS2024 program[85–90] on a finite-sized, boat-shaped, Si nanocrystal model ($Si_{308}H_{246}$) with a 3x14 unit cell surface and otherwise H-capped all around. Calculations were performed with the GGA(PBE) exchange-correlation functional.[91,92] This DFT code uses localized basis-functions, of either Slater-type orbital (STO) or numerical atomic orbital (NAO) type, together with frozen core approximations of the inner electron shells of the atoms in the system. In our case, a double-zeta basis set was used for all the atoms, except for a region consisting of the clean dimer atoms and the 4 Si atoms directly bonded to it, for which a double-zeta polarized basis set was used. This latter region was allowed flexibility during geometry optimization, while the other atoms were fixed. Note that the ground state is doubly degenerate. For each ground state geometry found, there is another configuration with the same energy and opposite dimer bond orientation.

For better accuracy of the final electronic structure, the optimized geometries were then transferred (the coordinates of the atoms belonging to the wire) to a periodic slab model, in order to perform single point calculations with higher numerical quality. For the periodic model calculations, we used a $Si_{672}H_{228}$ supercell with lateral extent of 3x14 surface unit cells and 8 Si layers in depth. The top and the bottom of the slab were hydrogen capped. For an efficient calculation, only the gamma point was used for k-space and the supercell was divided in two regions to be treated with either a coarse or a fine accuracy, according to their importance in the present study. The first or "central" region consists of the atoms of the dimer row where the dimer wire is located, minus the end Si dimers adjacent to the border of the supercell. The remaining atoms were allocated to a "support" region. A double-zeta polarized basis set was used for the atoms in the central region, while just a double-zeta basis for the support region. We made use of a frozen-core approximation for all atoms. The BAND DFT engine of AMS software was used for all calculations presented.

**4.7 Electronic Transport Calculations** We present theoretical predictions of quantum transmission using the NEGF-DFT method. We study short wires at the surface of finite-sized silicon nanoclusters having a Si(100) 2x1 surface reconstruction and hydrogen termination on all other sides. Each such cluster was sandwiched between two silver leads symmetrically placed to make contact with wire ends. These calculations were performed using the Amsterdam Modeling Suite[85–90] with the BAND computation engine. For computation efficiency, an approximation was applied for all transport calculations.[93] The self-consistent NEGF-DFT calculation cycle is applied only to a region usually dubbed the "extended molecule", in our case composed of the silicon cluster (containing the dimer wires) and the two finite-sized metal leads, contacting the cluster (and our wire) on each end. The self-energy of the semi-infinite electrodes is added post-SCF convergence in order to extract transport properties. Effectively, this approximation limits the applicability to a transport regime close to equilibrium (around zero bias), but in that regime, the accuracy is as good as for the fully self-consistent NEGF method.

For each transport calculation, the geometry of each dimer wire under study was extracted from a DFT geometry optimization calculation used also for the spectral studies. However, for transport, the model cluster was trimmed down to a minimum in order to make the calculation numerically feasible. The geometry was kept frozen during transport calculations. A double-zeta basis set was used in conjunction with frozen-core approximations. As electrode material, Ag was chosen for the range of work functions offered (4.26-4.74 eV), falling in the bulk silicon band gap and far from band edges, as well as for its known ability to adsorb on the silicon surface.[94,95] The (100) plane of the Ag crystal was interfaced with the Si crystal. Another consistency check was performed with respect to the number of Ag layers in the finite-sized leads in order to eliminate artifacts in the final results due to insufficient screening and decay of interactions within the far regions in the electrodes. It was found that 8 metallic layers were sufficient for that purpose.

**4.8 SiQAD DB Ground State Charge Simulations** The physical simulations for Figs. S1 and S2 were performed using *SimAnneal*, a simulation engine integrated into *SiQAD*[26] and accessible via *mnt.pyfiction* in Python. The surface parameters for the circuits in Figure S1 were $\lambda$=1.7 nm , $\mu$=-0.29 , and $\varepsilon$=4.1, which were chosen based on parameters extracted from refs [[5] and [26]]. The surface parameters for the circuits in Figure S2 were $\lambda$=5 nm , $\mu$=-0.31 , and $\varepsilon$=5.6, which were chosen based on parameters extracted from refs [26]. These values are known to shift with different crystals and sample preparation procedures.[26,61]

**Supporting Information**

Supporting Information including additional simulations, STM images, and spectroscopies is available to provide a more complete account of the entire dataset.

**Acknowledgements**

The authors would like to acknowledge funding from a Natural Sciences and Engineering Research Council of Canada: Alliance Quantum Grant to advance quantum sensing technologies.

**Authors' Contributions**

All fabrication and measurements were done by Max Yuan. The experiment was designed by Max Yuan, Roshan Achal, and Robert Wolkow. The simulations were designed and carried out by Lucian Livadaru. The text was primarily written by Max Yuan, Lucian Livadaru, and Robert Wolkow. All authors were involved with data analysis and manuscript editing.

**Conflicts of Interest**

Quantum Silicon Inc. is seeking to commercialize atomic silicon quantum dot-based technologies. The remaining authors declare no competing interests.

**Supporting Information: Engineering Quantum Wire States for Atom Scale Circuitry**

Max Yuan[1], Lucian Livadaru[1], Roshan Achal[2], Jason Pitters[3], Furkan Altincicek[1], Robert Wolkow[1,2]

[1]Department of Physics, University of Alberta, Edmonton, Alberta, T6G 2E1, Canada
[2]Quantum Silicon Inc., Edmonton, Alberta, T6G 2M9, Canada
[3]National Research Council of Canada, Edmonton, Alberta, T6G 2M9, Canada

Corresponding email: rwolkow@ualberta.ca

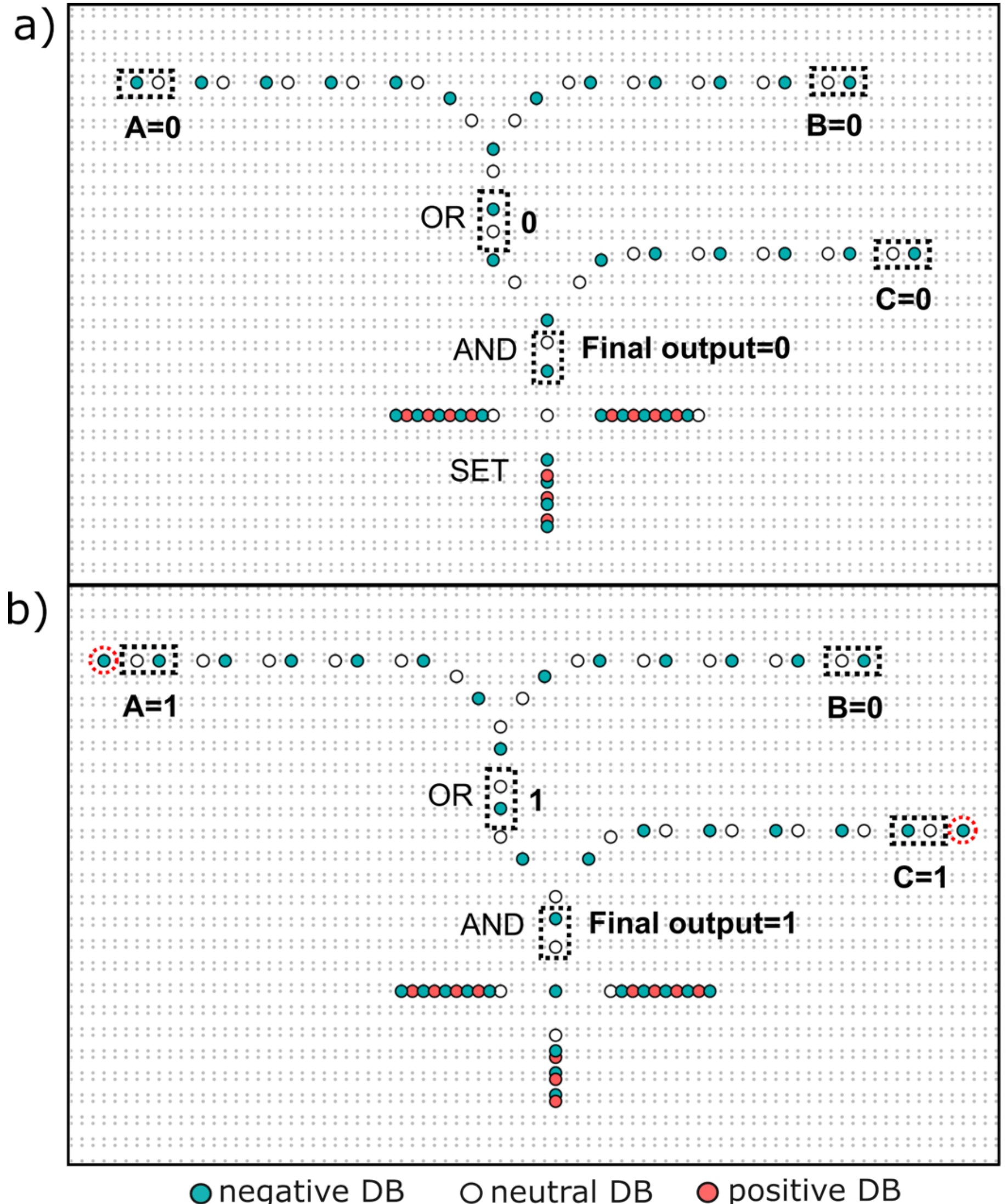

**Fig. S1a** SiQAD [26] ground state charge simulation results for the prototype DB logic gate circuit from Fig. 1a using the binary atomic silicon logic circuitry scheme from ref [5]. The simulations were done using SimAnneal with λ=1.7 nm , μ=-0.29 , and ε=4.1, which were chosen based on parameters extracted from refs [5 and 26]. This circuit combines an OR gate, an AND gate, and a SET. Here, pairs of neutral and negative DBs make up binary bits, with the bit state (0 or 1) dependent on the position of the negative DB within the pair. The binary information at the inputs (A, B, and C) then propagates through the logic gate structures, replicating the correct truth table. In this case all inputs are set to 0 resulting in an output of 0. The DB SET is positioned at the bottom to determine the output value by detecting the charge of the output DB. **b** The same circuit as (a) but with altered input values. Here, the input DBs outlined in red switch inputs A and C into the binary 1 state. The final output then changes to a 1 in agreement with the expected result for this circuit.

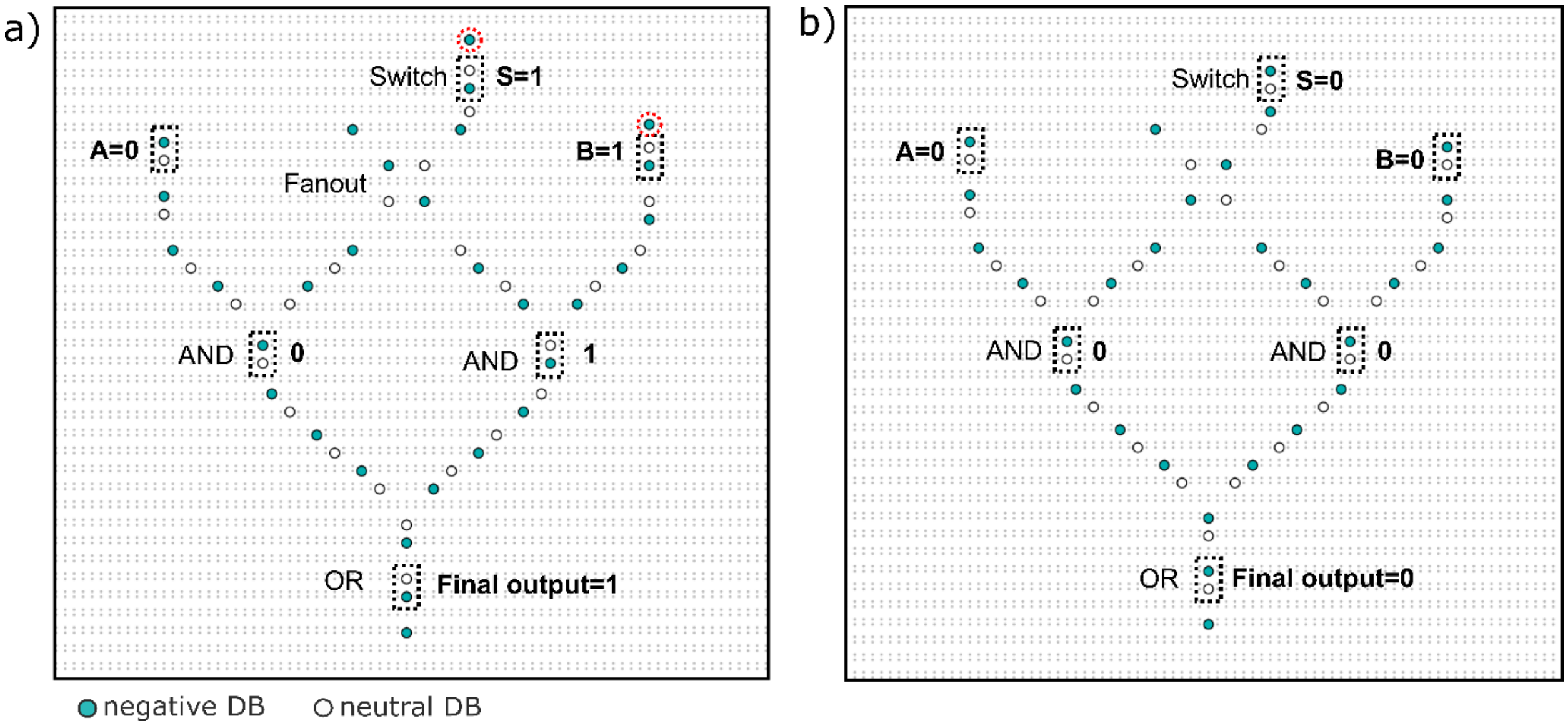

**Fig. S2a** SiQAD ground state charge simulation results for the prototype DB multiplexing circuit from Fig. 1b using the same binary atomic silicon logic circuitry scheme from ref [5]. The simulations were done using SimAnneal with λ=5 nm , μ=-0.31 , and ε=5.6, which were chosen based on parameters used in ref [26]. This circuit combines a switch, a fanout, two AND gates, and an OR gate. Similar to Fig. S1, pairs of neutral and negative DBs make up binary bits, with the bit state (0 or 1) dependent on the position of the negative DB within the pair. Here, input DBs outlined in red set the switch and input B into the binary 1 state, resulting in a final output of 1, as expected for a multiplexer. **b** The same circuit as (a) but without input DBs, resulting in inputs being set to 0 and a final output of 0.

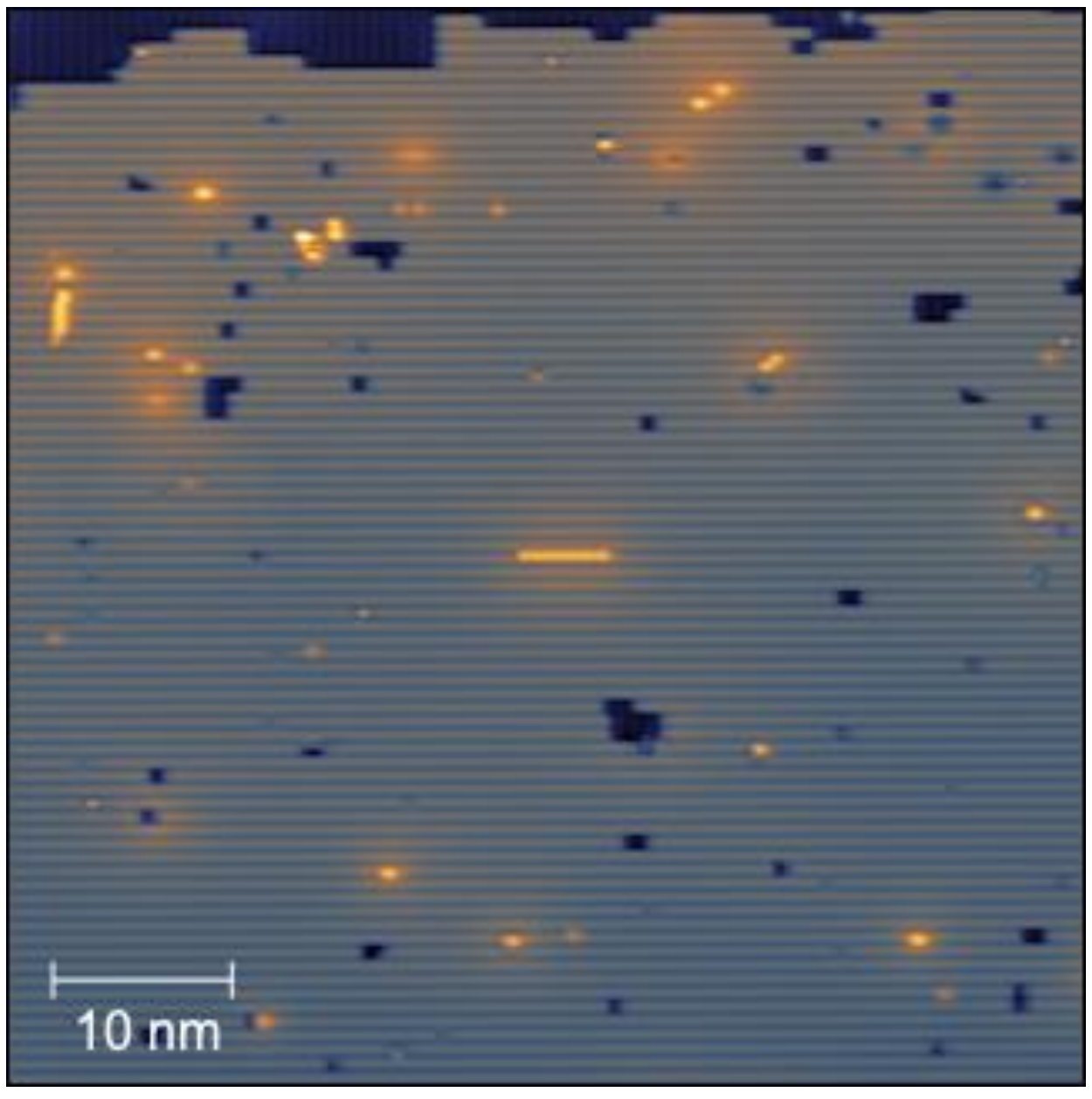

**Fig. S3** STM image (-1.8 V 50 pA) centered on the clean 15 nm x 25 nm defect free area where the wires were fabricated.

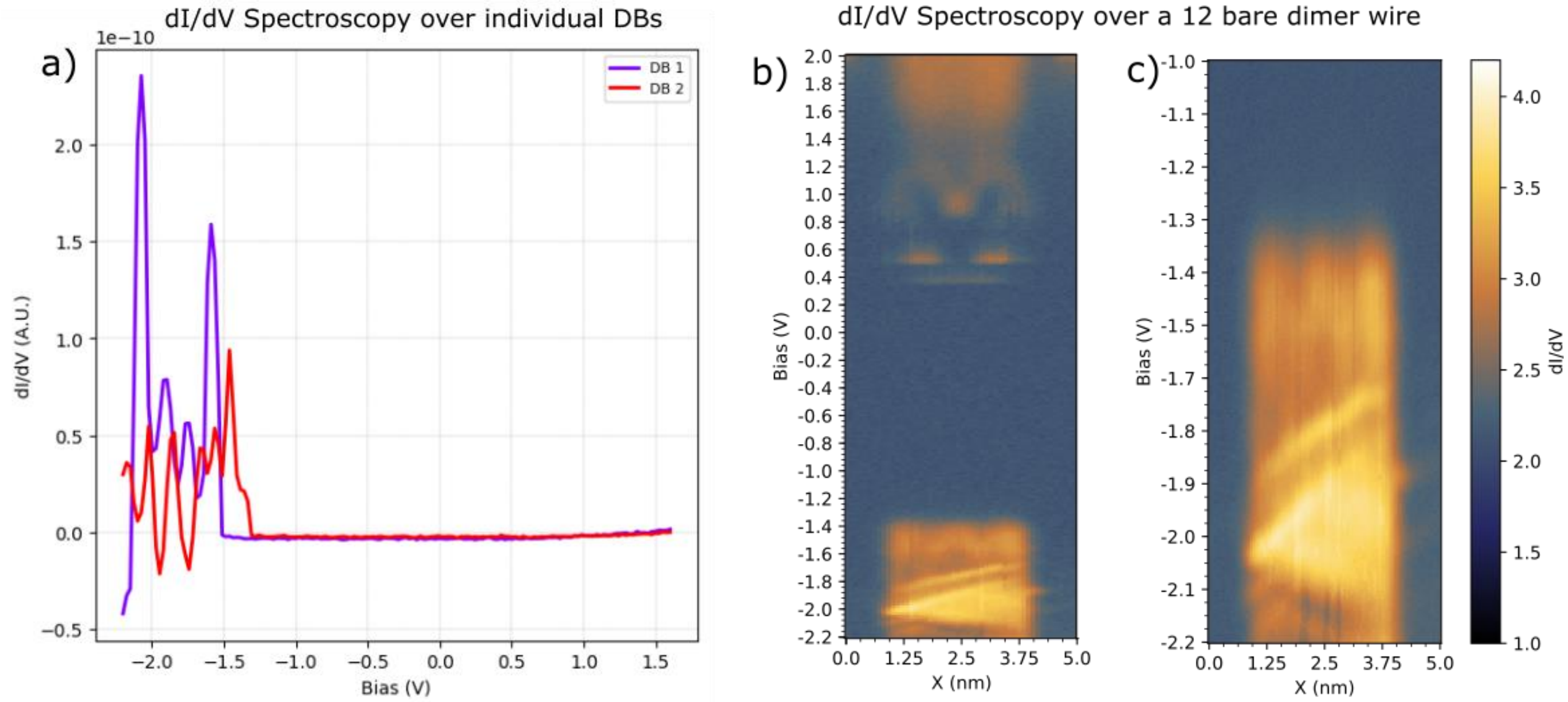

**Fig. S4a** dI/dV spectroscopies done over two different isolated DBs. Spectral peaks at negative biases obscure the DB features and arise due to nearby dopants and charge centers. The curve in Fig 3b does not show these same features due to its distance from dopants and the dopant depletion region. **b** and **c** dI/dV line spectroscopies done by combining 100 individual spectra over a 10 bare dimer long wire with nearby charge defects/dopants. Again, the negative wire states are concealed by the dopant related features.

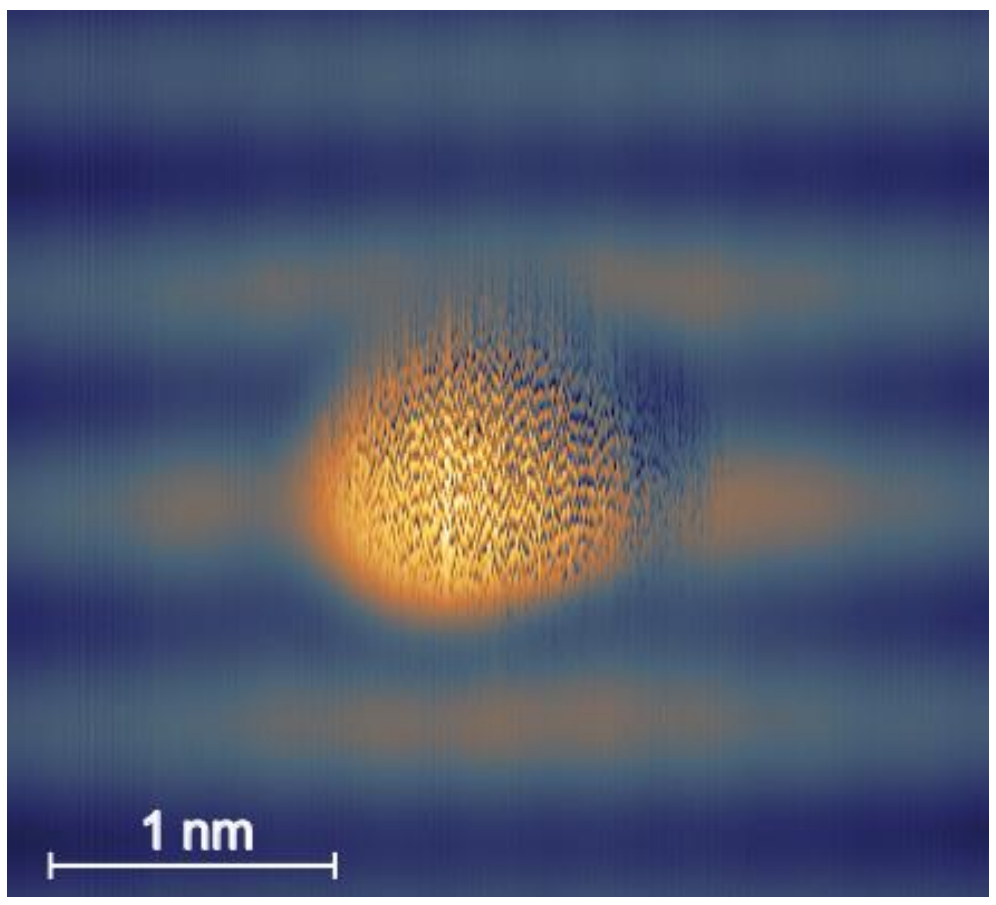

**Fig. S5** STM image (-1.5 V 50 pA) of a DB that is on average neutrally charged. At this bias, the filling and emptying rates through the DB are approximately equivalent causing the streaky, fluctuating charge related appearance of this DB. This tip bias coincides with the -1.5 V dI/dV peak in Fig. 3b.

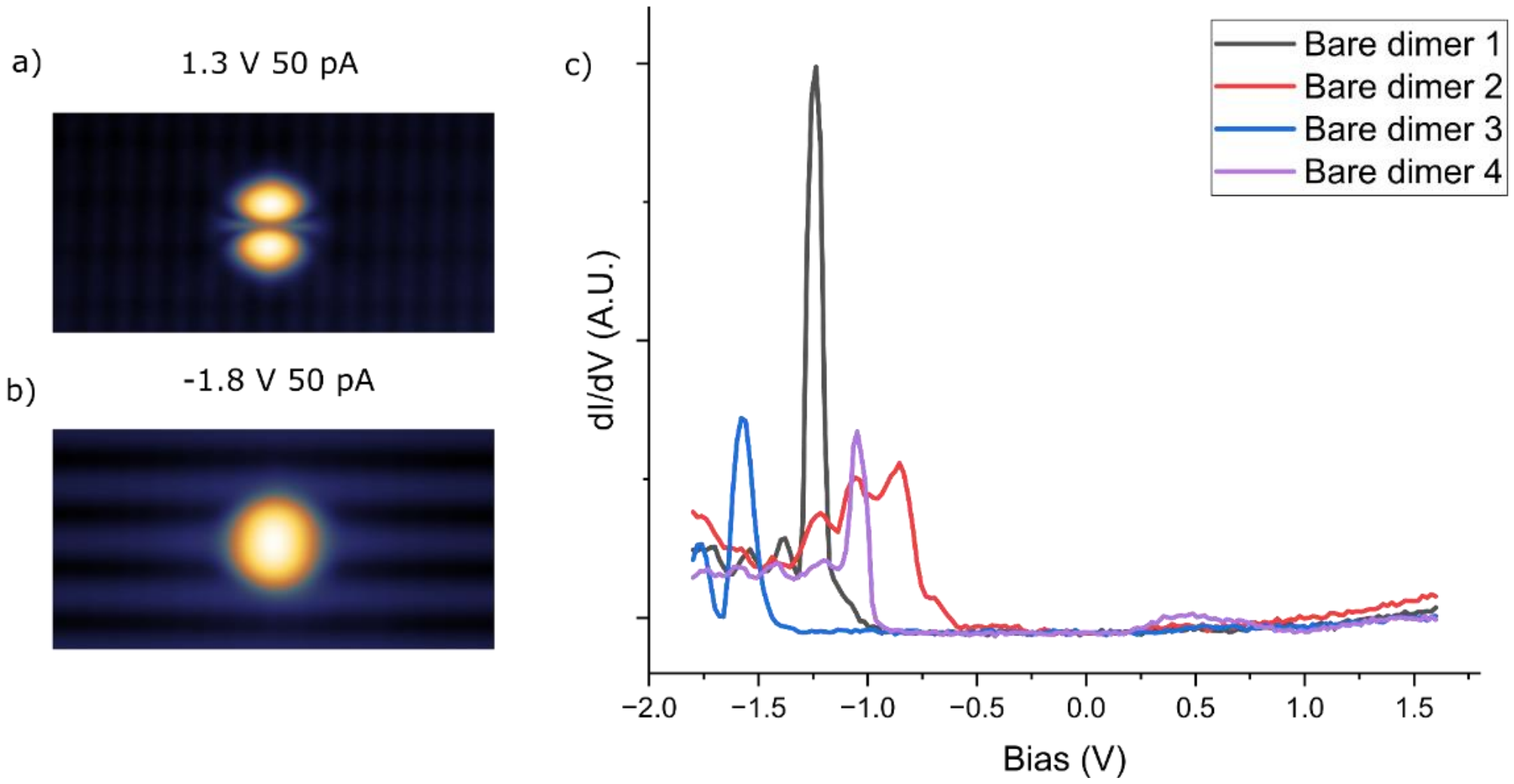

**Fig. S6a and b** Empty and filled state, STM images of a bare dimer. **c** dI/dV spectroscopies done over the center of four different bare dimers. The different spatial locations of each bare dimer introduce electrostatic differences which distort the bare dimer electronic signature and result in significantly different spectra. A common tip setpoint of 1.8 V 20 pA over H-Si was used for each measurement.

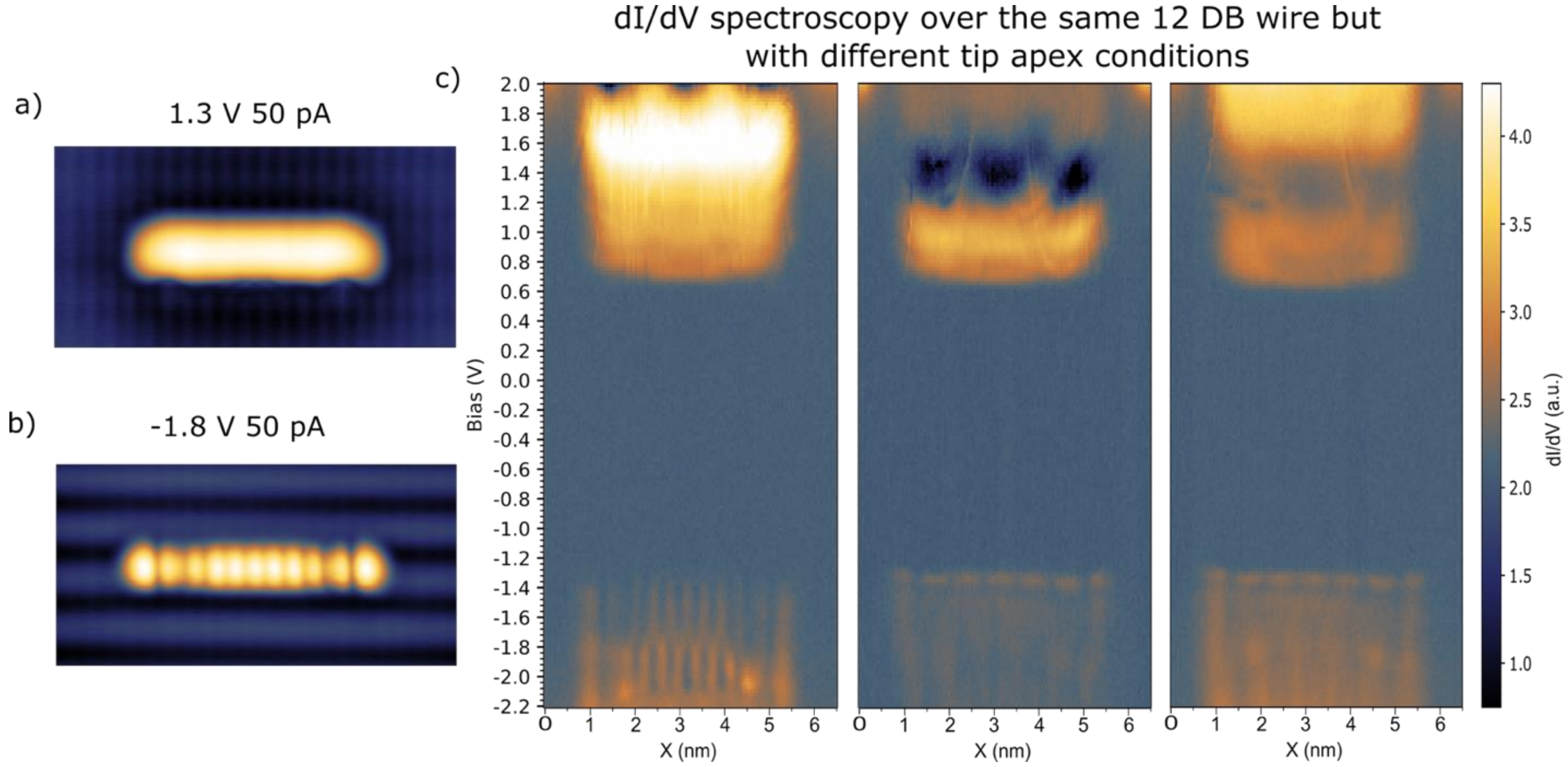

**Fig. S7a and b** Empty and filled state, STM images of a 12 DB single atom wire. **c** Three dI/dV line spectroscopies done over the same 12 DB single atom wire. Differences in tip apex conditions result in significantly different spectra, showing the importance of maintaining consistent tip conditions when comparing spectroscopies. Each measurement was done with the same 1.8 V 30 pA tip setpoint.

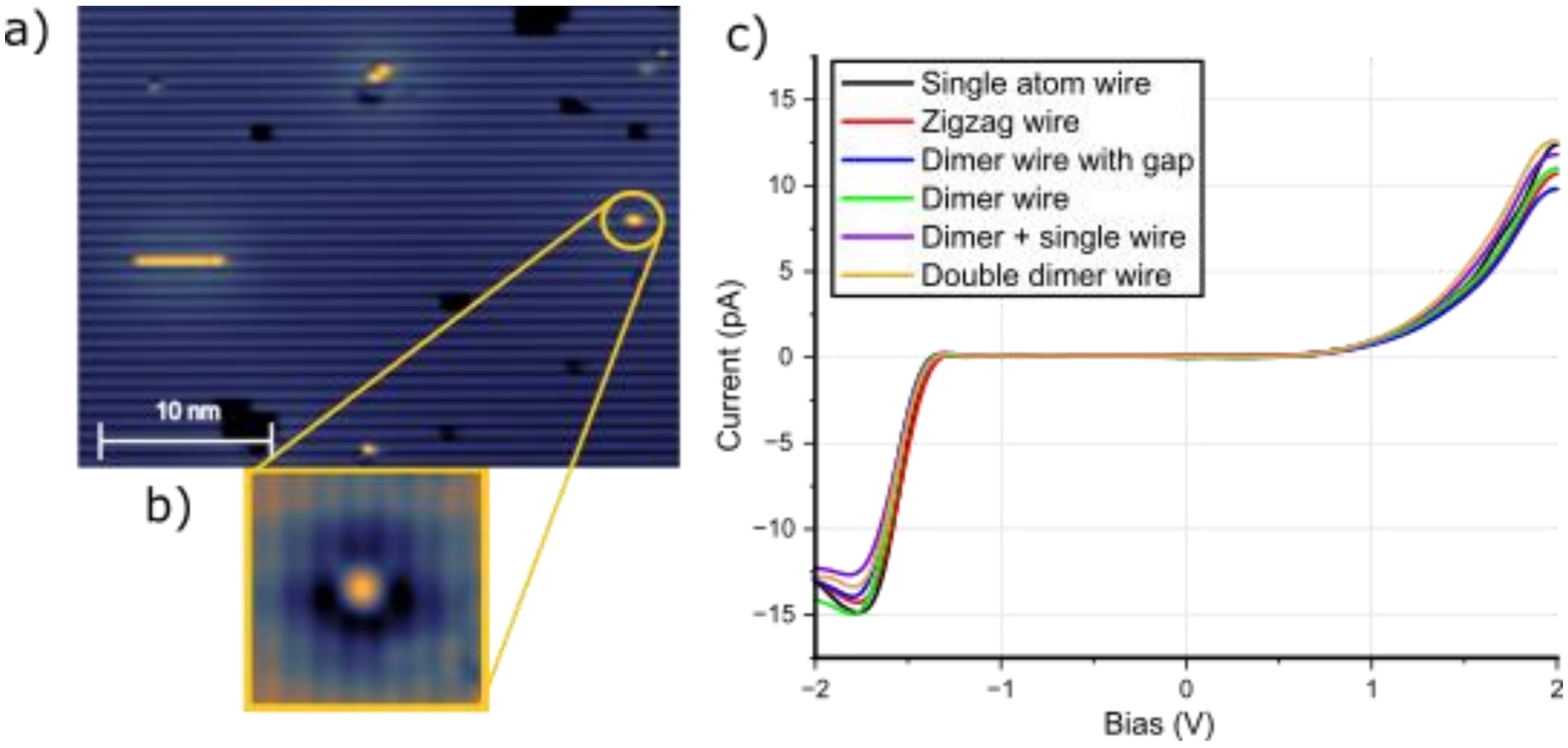

**Fig. S8a** STM image (-1.8 V 50 pA) showing the experimental area with the reference DB located 25 nm away from the wire. **b** STM image (1.3 V 50 pA) of the reference DB. **c** IV spectroscopy done over the reference DB before analysis of each wire. Each spectrum shows the same characteristic curve indicating similar tip conditions. Slight variability was allowed to account for small deviations in tip position. When a tip change occurred, tip shaping was done until this curve could be replicated.

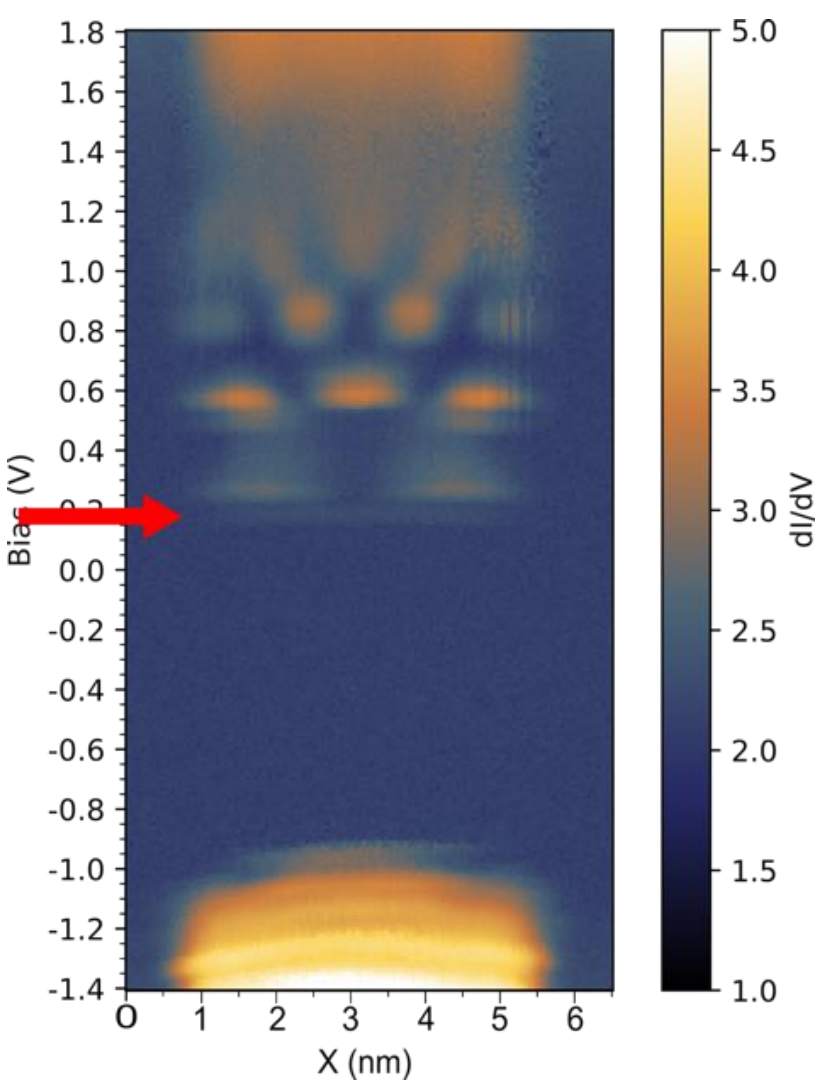

**Fig. S9** A dI/dV line spectroscopy done over a 12 dimer long wire, similar to the measurement in Fig. 4. Notably, the tip apex condition during this measurement allowed for the first LUMO state to be resolved at around 110 mV as indicated by the red arrow. The result in the main text was not able to see this due to very low signal and having a different tip apex at the time.

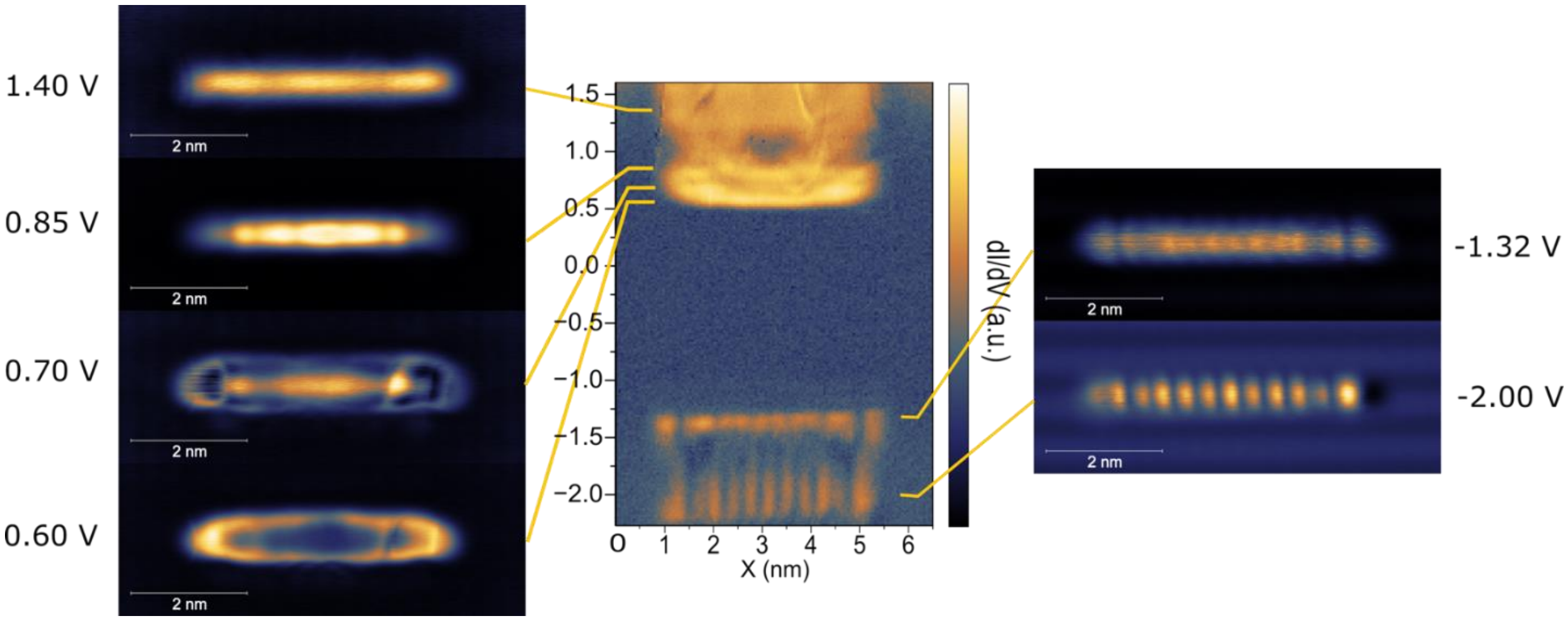

**Fig. S10** Constant height dI/dV images taken at energies chosen from the dI/dV line spectroscopy over the 12 DB single atom wire (Fig. 2c and Fig. 5a). The positive bias images were taken from a tip setpoint of 1.3 V 50 pA over H-Si, and the negative bias images were taken from a tip setpoint of -1.8 V 50 pA over H-Si.

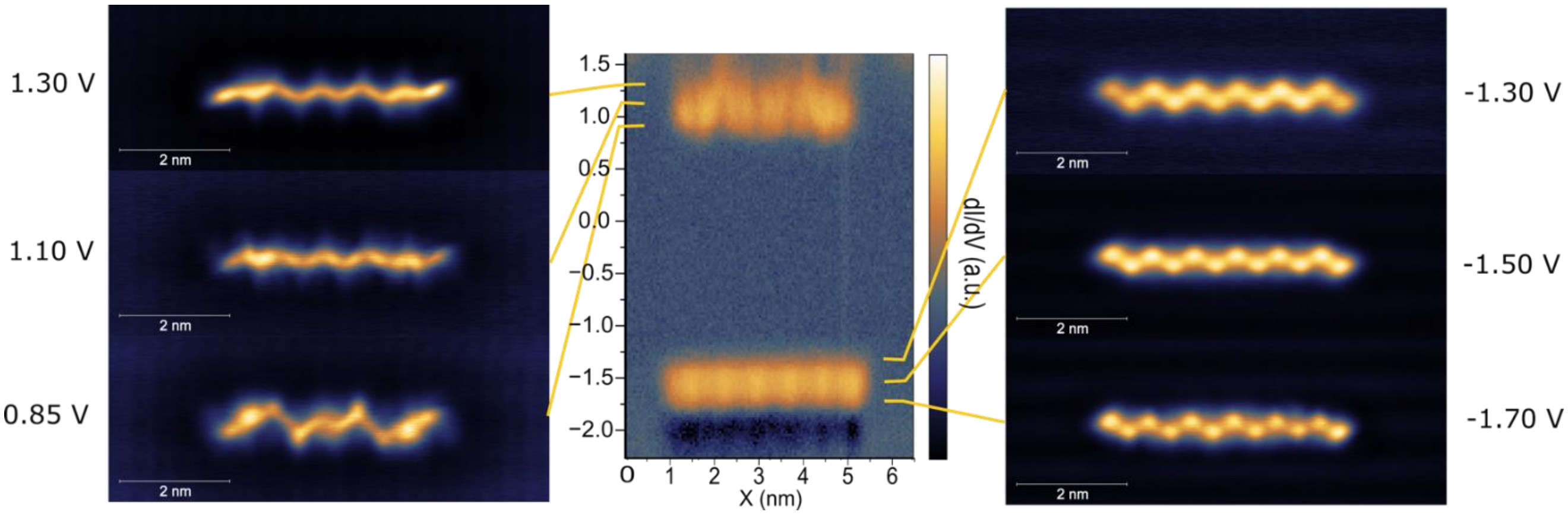

**Fig. S11** Constant height dI/dV images taken at energies chosen from the dI/dV line spectroscopy over the 12 long zigzag wire (Fig. 2d and Fig. 5b). The positive bias images were taken from a tip setpoint of 1.3 V 50 pA over H-Si, and the negative bias images were taken from a tip setpoint of -1.8 V 50 pA over H-Si.

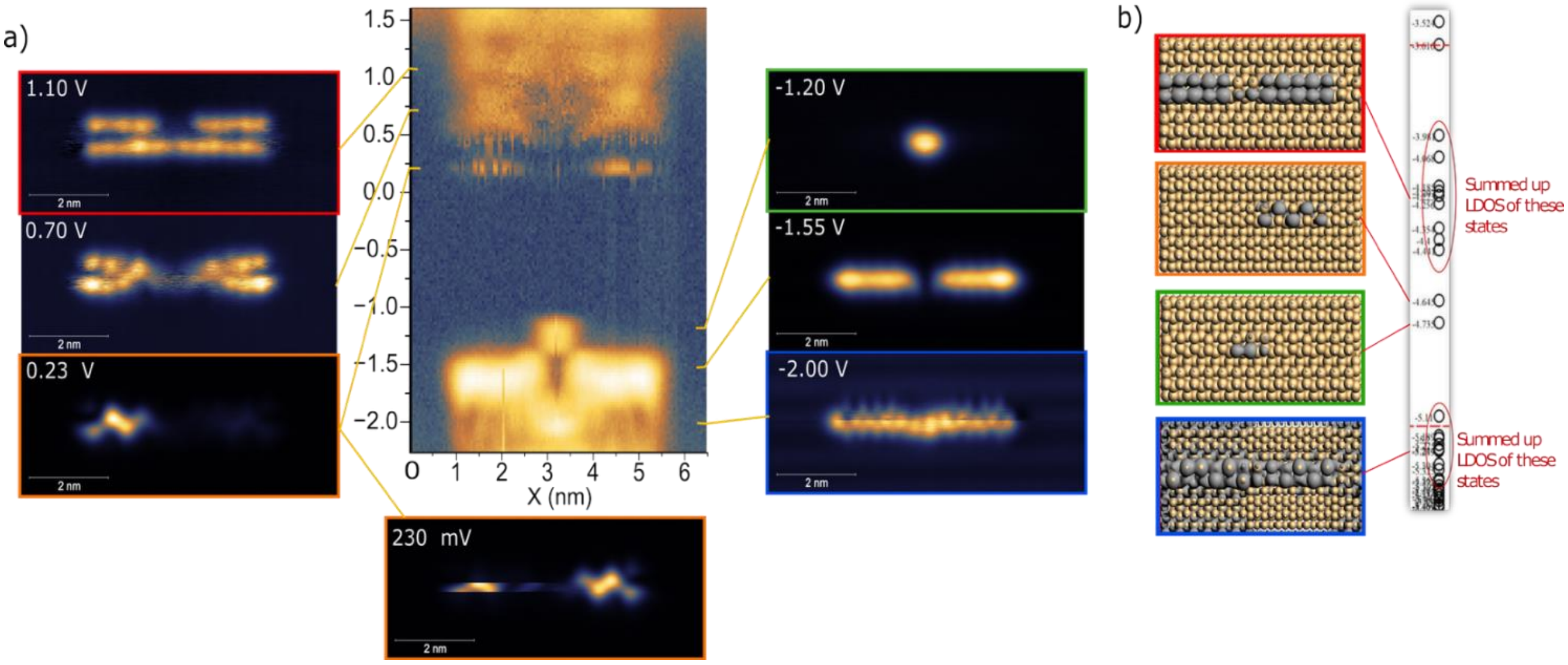

**Fig. S12a** Constant height dI/dV images taken at energies chosen from the dI/dV line spectroscopy over the 12 long dimer wire with a 2H gap (Fig. 1e and Fig. 5c). The positive bias images were taken from a tip setpoint of 1.3 V 50 pA over H-Si, and the negative bias images were taken from a tip setpoint of -1.8 V 50 pA over H-Si. The LUMO state at 239 mV shows switching between two degenerate configurations. **b** DFT based LDOS calculations for the dimer wire with the gap. The simulated state wavefunctions closely resemble their respective dI/dV images.

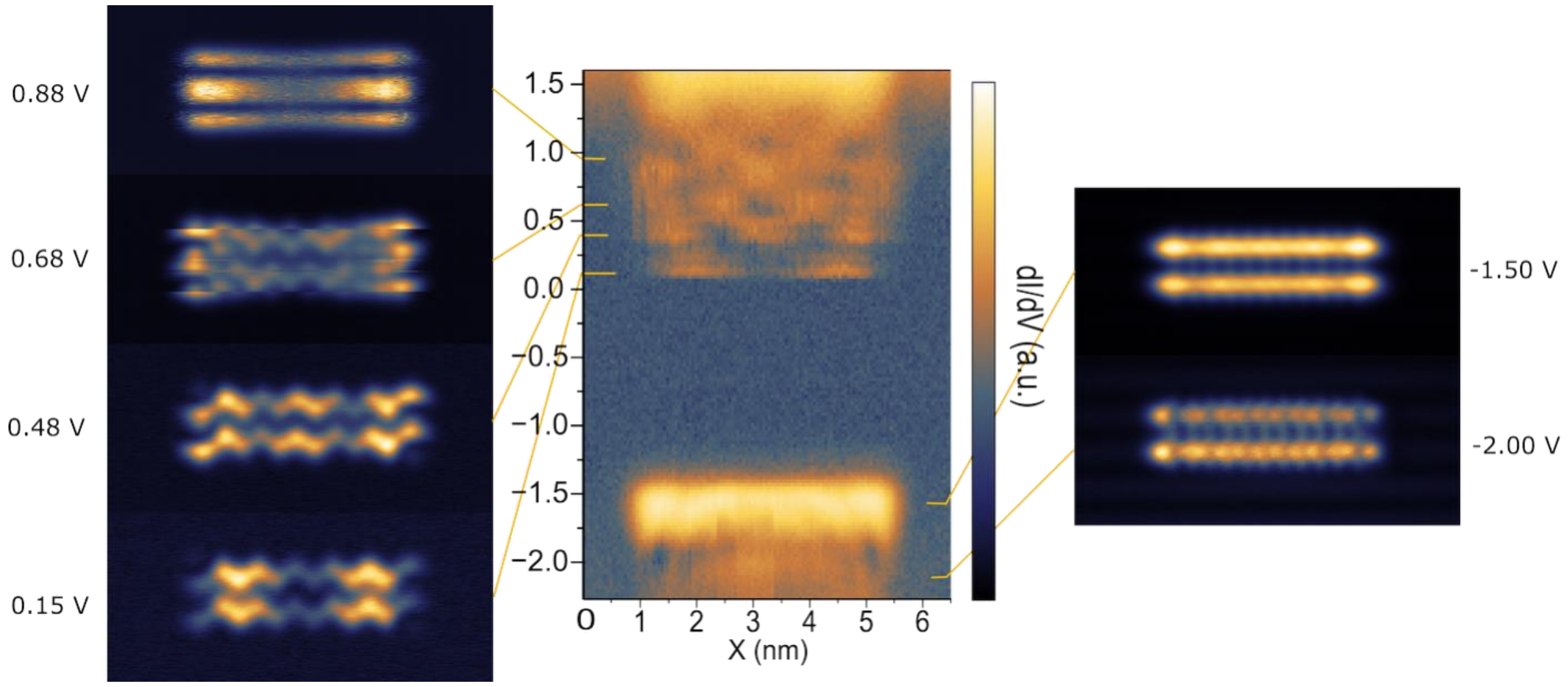

**Fig. S13** Constant height dI/dV images taken at energies chosen from the dI/dV line spectroscopy over the 12 long double dimer wire (Fig. 2h and Fig. 5e). The positive bias images were taken from a tip setpoint of 1.3 V 50 pA over H-Si, and the negative bias images were taken from a tip setpoint of -1.8 V 50 pA over H-Si.

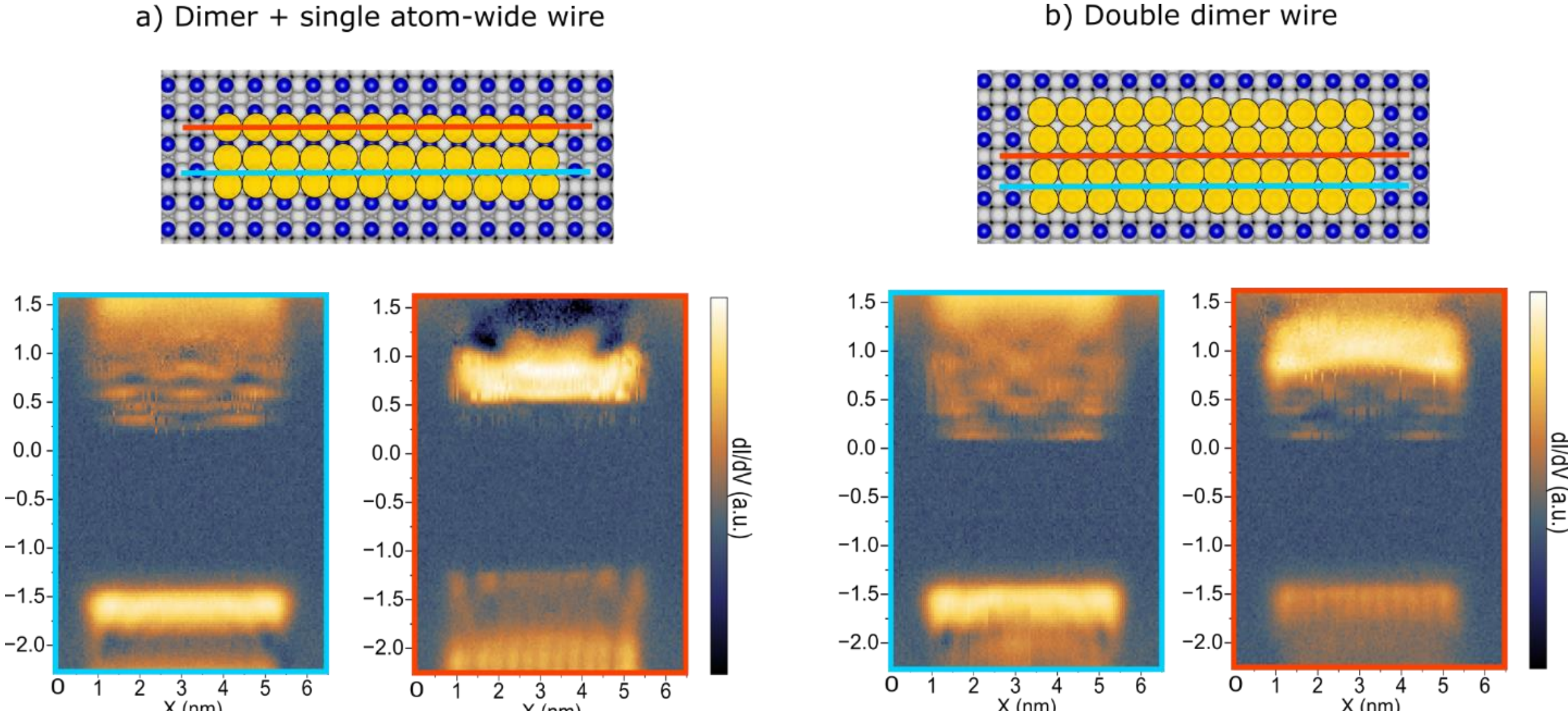

**Fig. S14a and b** dI/dV spectroscopies done over the dimer plus single atom-wide wire and the double dimer wire. The blue lines in the ball and stick diagrams show the position of the spectroscopies shown in Fig. 5d and e of the main text. The red lines show the position of the alternate spectroscopies which are displayed here. The respective spectroscopies are represented by the color of their border. For the double dimer wire in b, the fact the same states are visible over the dimer and in between the dimer rows indicates that transmission across rows may be feasible.

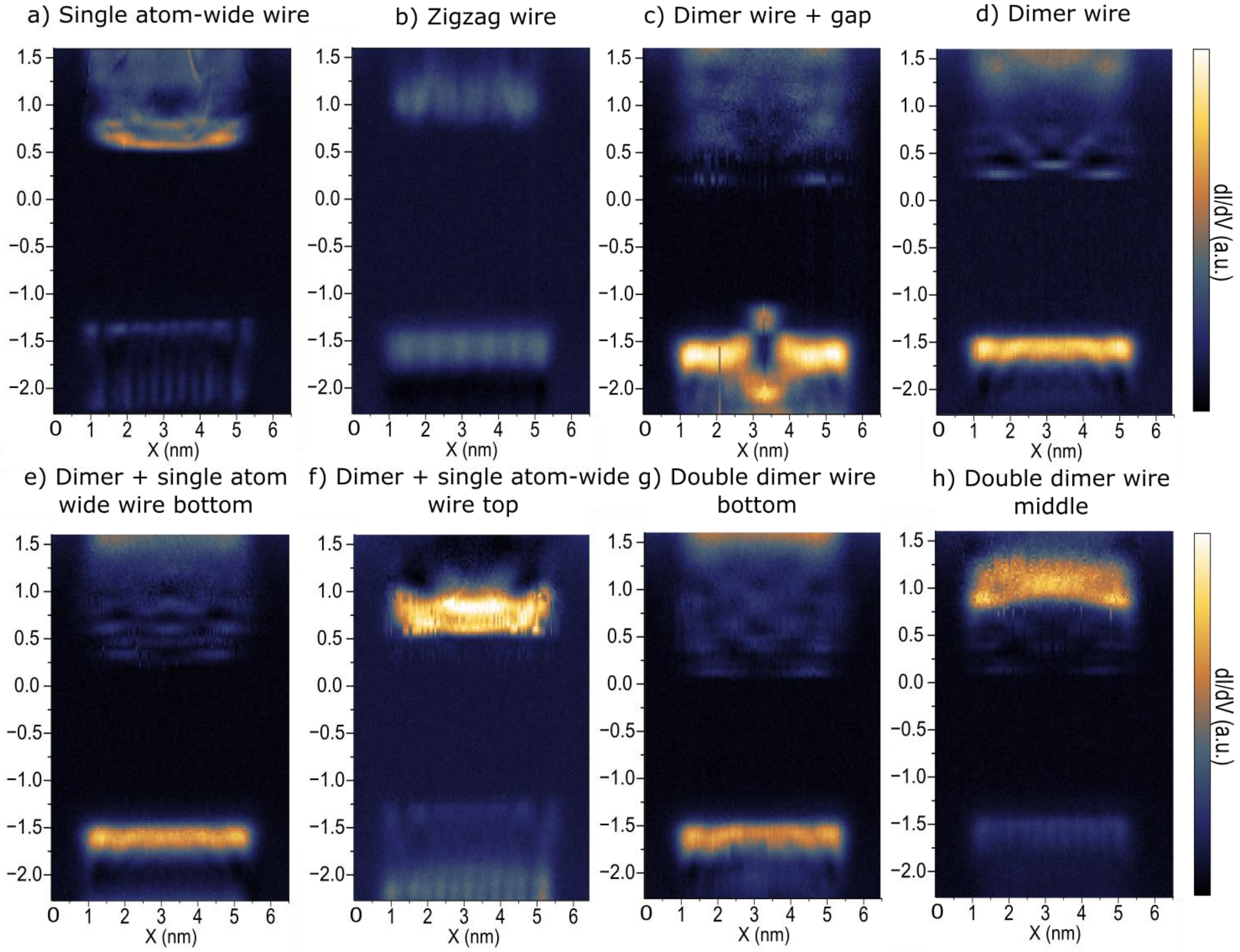

**Fig. S15a-h** The dI/dV spectroscopies without application of logarithm to enhance visibility corresponding to the spectra in Figs. 4, 5 and S12. The LDOS maps were log processed in the main text to increase the visibility of lower magnitude states.

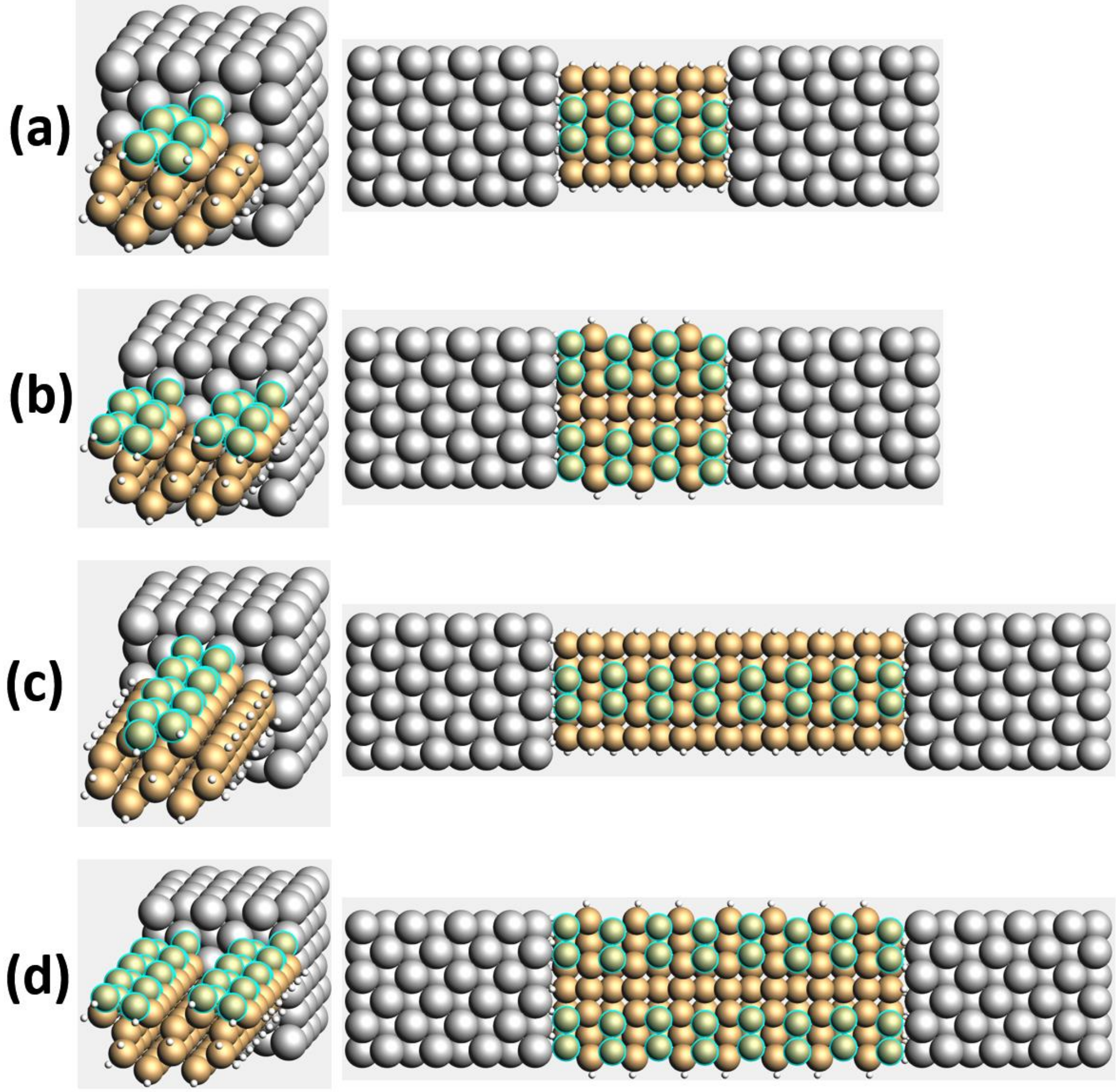

**Fig. S16** Molecular models showing the system used for simulating quantum transport properties by NEGF-DFT. The ground-state geometries used in the calculations of four wire types: **a** 4-long single dimer wire; **b** 4-long double dimer wire; **c** 8-long single-dimer wire; **d** 8-long double dimer wire. Left side: perspective view with near-side electrode removed. Right side: top view with both electrodes shown. Silicon atoms are depicted in yellow, H atoms in white, Ag atoms in gray, and the Si atoms hosting DBs are highlighted in light green.